\documentclass[aps,pra,twoside]{revtex4}
\usepackage{graphicx}
\usepackage{epsfig}
\usepackage{amsmath,amssymb,revsymb,latexsym}

\newcommand{\ID}{{\operatorname{id}}}
\def\duzomniejsze{<\kern-.7mm<}
\def\duzowieksze{>\kern-.7mm>}

\def\textbf#1{{\bf #1}}
\def\beq{\begin{equation}}
\def\eeq{\end{equation}}
\def\be{\begin{equation}}
\def\ee{\end{equation}}
\def\ben{\begin{eqnarray}}
\def\een{\end{eqnarray}}
\def\beqa{\begin{eqnarray}}
\def\eeqa{\end{eqnarray}}
\def\eea{\end{array}}
\def\bea{\begin{array}}
\newcommand{\bei}{\begin{itemize}}
\newcommand{\eei}{\end{itemize}}
\newcommand{\bee}{\begin{enumerate}}
\newcommand{\eee}{\end{enumerate}}
\newcommand{\merging}{{\langle {\rm id}_{A\rightarrow B'} : \rho_{AB} \rangle}}

\def\tr{{\operatorname{Tr}\,}}

\def\ra{\rangle}

\def\>{\rangle}
\def\<{\langle}

\def\ot{\otimes}

\def\dt#1{{{\kern -.0mm\rm d}}#1\,}

\newcommand{\tA}{{\widetilde{A}}}
\newcommand{\tB}{{\widetilde{B}}}
\newcommand{\tR}{{\widetilde{R}}}

\def\sigalpe{{\sigma_\alpha'}^{\kern-.7mm E}}
\def\sigalpb{{\sigma_\alpha'}^{\kern-.7mm B}}

\def\phiprimj{{{\phi_{ABR}'}^{\kern-6mm(j)}\kern2mm}}
\def\phiprimjout{{{\phi_{ABB'B''R}'}^{\kern-12mm(j)}\kern9mm}}

\def\phijout
{\phi_{ABR}^{j\ out}}

\def\rhoprimj{{ {\rho'_{AR}}^{\kern-3.5mm(j)} \kern2mm}}

\def\phiprimi{{{\phi_{ABR}'}^{\kern-6mm(i)}\kern2mm}}
\def\phiid{{{\phi^{id}_{ABR}}^{\kern-3mm(j)}\kern2mm}}

\newcommand{\rk}{{\operatorname{rank}\,}}

\def\ep{\epsilon}

\def\lavv{\biggl\<}
\def\ravv{\biggl\>}

\def \eass {E_A}

\def \sab {S(AB)}
\def \sa {S(A)}
\def \sb {S(B)}
\def \sacb {S(A|B)}
\def \sbca {S(B|A)}

\def \ra {\rangle}

\def \r {\rho}

\def \ra {\rho_A}

\def \eass {E_A}
\newcommand{\proj}[1]{\ket{#1}\!\bra{#1}}
\newcommand{\bra}[1]{\langle #1 |}
\newcommand{\ket}[1]{| #1 \rangle}

\newtheorem{lemma}{Lemma}

\newtheorem{theorem}[lemma]{Theorem}
\newtheorem{proposition}[lemma]{Proposition}
\newtheorem{definition}[lemma]{Definition}
\newtheorem{expl}[lemma]{Example}
\newtheorem{remark}[lemma]{Remark}
\newtheorem{conjecture}[lemma]{Conjecture}

\newlength{\blank}
\newenvironment{beweis}[1][{\hspace{-\blank}}]{{\noindent\emph{Proof~{#1}.\ }}}{\hfill$\Box$\vskip 0.5\baselineskip}

\begin{document}

\title{Quantum state merging and negative information}
\author{Micha\l{} Horodecki}
\affiliation{Institute of Theoretical Physics and Astrophysics, University of Gda\'nsk, 80-952 Gda\'nsk, Poland}

\author{Jonathan Oppenheim}
\affiliation{Department of Applied Mathematics and Theoretical Physics, University of Cambridge, Cambridge CB3 0WA, U.K.}

\author{Andreas Winter}
\affiliation{Department of Mathematics, University of Bristol, Bristol BS8 1TW, U.K.}

\begin{abstract}
We consider a quantum state shared between many distant locations,
and define a quantum information processing primitive, \emph{state merging},
that optimally {\it merges} the state into one location.
As announced in [Horodecki, Oppenheim, Winter, Nature {\bf 436}, 673 (2005)],
the optimal entanglement cost of this task is the conditional entropy
if classical communication is free.  Since this quantity can be negative, and 
the state merging rate measures partial quantum information, we find that quantum information
can be negative.
The classical communication rate also has a minimum rate:
a certain quantum mutual information. State merging enabled one to solve a number of open problems: distributed
quantum data compression, quantum coding with side information at the decoder and sender,
multi-party entanglement of assistance, and the capacity of the quantum multiple
access channel.   It also provides an operational proof
of strong subadditivity.
Here, we give precise definitions and prove these results rigorously.
\end{abstract}

\date{26th December 2005}

\maketitle

\section{Introduction}
\label{sec:intro}
The field of quantum information theory is still in its infancy, with many of the
key building blocks of the theory not yet in place or not well understood.
This is perhaps not surprising, since the important elements of
classical information theory have only been in place since the 70's.
The notion of classical information was first introduced by Shannon~\cite{Shannon1948} 
who defined it operationally, as the minimum number of bits needed to communicate
the message produced by a statistical source.   
This gave meaning to the entropy $H(X)$ of
the source  producing a random variable $X$.  
The amount of information that two random variables $X$ and $Y$ have in common
was given a meaning through the mutual information $I(X:Y)=H(X)+H(Y)-H(XY)$.  Operationally
it is the rate of communication possible through a noisy channel taking $X$ to $Y$. 
The fundamental Shannon theorems treated two basic questions:
how many bits does one need to transmit a message from a source?
How many bits can one send via a noisy channel?

Another basic brick in classical information theory, 
which is a generalization of the noiseless coding problem,
is the notion of partial information. The question is now, how many bits 
does the sender (Alice) need to send to transmit a message from the source, provided 
the receiver (Bob) already has some prior information about the source. 
The amount of bits we call the partial information. Slepian and Wolf 
showed that partial information is equal to the entropy of the source reduced by 
the mutual information~\cite{slepian-wolf}. 
This quantity is equal to what is called conditional entropy $H(X|Y)=H(XY)-H(Y)$. 
It is actually an entropy, and was originally defined as the average entropy of conditional 
probability distributions:
\be
H(X|Y)=-\sum_{xy} p_Y(y) p_{X|Y}(x|y)\log p_{X|Y}(x|y),
\ee
with $p_{X|Y}(x|y)$ the probability of the source producing symbol $x$ conditioned on
the fact that Bob has $y$, and $p_Y(y)$ the probability that $y$ is produced at Bob's site.

\begin{figure}
\label{fig:classicalinfo}
\centering
\includegraphics[scale=1.0]{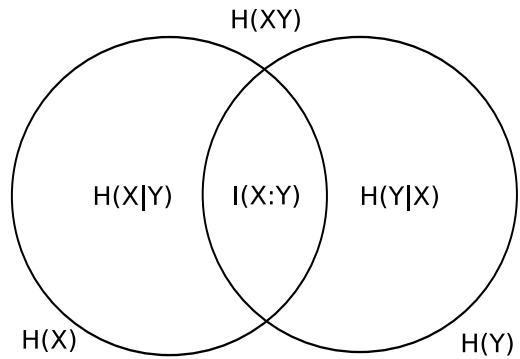}
\caption{A graphical representation of the building blocks
of classical information theory.  The total information of the
source producing pairs of random variables $X,Y$ is $H(XY)$,
while the information contained in just variable $X$ ($Y$)
is $H(X)$ ($H(Y)$).  The information common to both variables
is the mutual information $I(X:Y)$, while the partial informations
are $H(X|Y)$ and $H(Y|X)$.  In the quantum case, the 
quantum mutual information $I(A:B)$ can be greater than the total
information $S(AB)$, which can be also greater than the 
local informations $S(A)$ and $S(B)$.  To compensate, the
partial informations $S(A|B)$ and $S(B|A)$ can be negative.}
\end{figure}

This discovery of Slepian and Wolf clarified the picture of correlated 
sources: mutual information is the knowledge common to both Alice and  Bob.
Entropy of Alice's source is its full information content.
The difference between the two is the information that Bobs needs to 
complete his prior knowledge about Alice's source (Figure \ref{fig:classicalinfo}).  
It thus provided an information
theoretic basis for the conditional entropy.  It should be noted that it is a
highly non-trivial operation, since Alice is able to communicate to Bob the
full information about her string $X_1\ldots X_n$, even though she is
unaware of what string $Y_1\ldots Y_n$ Bob has.
\begin{table}
\label{tab:classical}
\begin{tabular}{|l|l|p{6.5cm}|}
	\hline
\textbf{\em concept}  &  \textbf{\em quantity}     &   \textbf{\em operational meaning}      \\
	\hline \hline
information   
& $H(X)$ 
& The rate at which a source can convey messages (Shannon compression) 
\\ \hline
mutual information   
& $I(X:Y)$ 
& For an input $X$ which produces $Y$ after being sent down a channel, $I(X:Y)$ is the rate
at which information can be sent reliably (channel coding)  
\\ \hline
partial information   
& $H(X|Y)$ 
& The rate at which messages $X$ can be sent to a party who has prior information $Y$ (Slepian-Wolf theorem)  \\
	\hline
\end{tabular}
\caption{Key concepts in classical information theory}
\end{table}

The quantities and operational meaning of the entropy, mutual information, and conditional
entropy thus form the basic building blocks of classical information theory. 
We are interested in finding the corresponding basic elements in quantum information 
theory.
The first step was done by Schumacher~\cite{Schumacher1995}, 
who showed that the von Neumann entropy plays an analogous role to Shannon 
entropy: it has the operational interpretation of the number of qubits needed 
to transmit quantum states emitted by a statistical source. 

The next step was to find an analogue of the noisy coding theorem.
Here it turned out that the analogy was not very strict: 
the quantum analogue of mutual information cannot be obtained by replacing 
Shannon entropies with von Neumann ones. It was found that 
the capacity of the quantum channel is determined by a different quantity -- the coherent 
information~\cite{Schumacher-Nielsen,BarnumNS-converse1997}.
The coherent information, defined for a bipartite state $\rho_{AB}$ is
\be
I(A\>B)= S(B) - S(AB),
\ee
and the channel capacity is obtained~\cite{Lloyd-cap,shor-cap,igor-cap} by maximising it 
over input states $\rho_A$. Here, $S(B)$ and $S(AB)$ are the von Neumann entropy of
states $\rho_B = \tr_A \rho_{AB}$ and $\rho_{AB}$,
and we adopt the notation of dropping the explicit 
dependence on $\rho$ when such dependence is obvious.

With the coherent information, there was a persistent mystery -- for any particular input
$\rho_A$, the quantity $S(A)-S(AB)$ could be negative, and it was not known how to interpret
such a quantity, as it indicated a sense in which the channel capacity could be negative for such
input distributions.  Thus it is often the case that for a particular channel, no inputs will
give positive distributions, and one should set the coherent information to zero,
by inputting the null distribution (any pure state).

Turning next to a quantum analogue of prior and partial information, there had previously not
been any such notion -- a quantum scenario like that of Slepian-Wolf
appeared intractable~\cite{adhw2004}. Another serious obstacle 
in the quantum world is that there are no conditional probabilities, hence 
conditional entropy cannot be defined. Conditional probabilities only exist after one performs
a measurement which of course destroys the state.  One may try to overcome 
this difficulty, by naively replacing Shannon entropies with von Neumann 
one in the formula for conditional entropy, 
so that quantum conditional entropy would be the difference between the total entropy 
and the entropy of subsystem.  
\be
S(A|B)=S(AB)-S(B)
\ee
Such an approach has been strongly advocated~\cite{cerfadami},
however while this \emph{$H$ goes to $S$} rule works for defining
information, it doesn't work for channel capacity, as  
mentioned above. It is thus not clear that it is the correct thing to do.
However there is more serious obstacle here: the conditional entropy
defined by taking $H$ to $S$ can be negative~\cite{Wehrl78,HH94-redun,cerfadami}.
In~\cite{HH94-redun} this problem was connected with quantum entanglement. Likewise
for maximally entangled states, it was connected with the ability to perform teleportation
\cite{cerfadami}.
It had already been noted by Schr\"odinger, that entangled state 
may possess a weird feature: if a system is in such a state 
we may know more about the whole system  than about subsystems. 
In~\cite{HH94-redun}, Schr\"odinger's intuition was quantified by von Neumann entropies, 
and it was found that the entropy of subsystem can be greater than 
the entropy of the total system only when the state is entangled. 
It was however also found that there are entangled states 
that do not exhibit this weird property. Thus there was a question:
what does it mean, that for some states we have such behaviour,
and not for other states?

It doesn't help that $-S(A|B)$ 
is nothing but the coherent information, that determines channel
capacity~\cite{Lloyd-cap,shor-cap,igor-cap}! How can the duality
between channel coding and Slepian-Wolf compression be conserved
in any quantum analog?

In our recent paper~\cite{SW-nature}, we approached the problem of quantifying
partial and prior information from a purely operational point of view.
Inspired by the classical Slepian-Wolf theorem, we consider the scenario in which an
unknown quantum state is distributed over two systems. We determined how much 
quantum communication is needed to transfer the full state to one system.  
This communication measures the \emph{partial information} one system needs
conditioned on its prior information. We found that the partial information
is given by the conditional entropy, just as in the classical case. 
However, in the classical case, partial information must always
be positive, while in the quantum world we find this physical
quantity can be negative.
If the partial information is positive, its sender
needs to communicate this number of quantum bits to the receiver to achieve state transfer;
if it is negative, the state can be transferred, and in addition, 
the sender and receiver \emph{gain} the corresponding potential for
future quantum communication.  This potential communication is in the form of pure entangled
states which can be used to teleport quantum states. Thus viewing entanglement
as a potential for quantum communication, we see that when the conditional entropy
is positive, entanglement needs to be consumed, while when it is negative,
entanglement is gained.

\begin{table}
\label{tab:quantum}
\begin{tabular}{|l|l|p{6.5cm}|}
	\hline
\textbf{\em concept}  &  \textbf{\em quantity}    &   \textbf{\em operational meaning}      \\
	\hline \hline
quantum information   
& $S(A)$ 
& The rate at which a source can convey quantum states (Schumacher compression) 
\\ \hline
coherent information   
& $I(A\rangle B)$ 
& For an input which produces $\rho_{AB}$ after being sent down a channel, $I(A\rangle B)$ is the rate
at which quantum information can be sent reliably down the channel (quantum channel coding). {\bf Merging allows
us to interpret the negative values of this quantity}  
\\ \hline
{\bf partial quantum information  } 
& $S(A|B)$ 
& {\bf The rate at which quantum states with density matrix 
$\rho_A$ can be sent to a party who has prior quantum information $\rho_B$ }  \\
	\hline
\end{tabular}
\caption{Key concepts in quantum information theory with additions due to merging highlighted in bold}
\end{table}

One can view it in another way -- the entropy $S(B)$ quantifies how
much Bob {\it knows} (in the sense of possessing the state), 
while the entropy $S(AB)$ quantifies how much there
is to know.  Since quantum distributions can have $S(AB)\leq S(B)$,
there is a sense in which Bob knows too much.  If Alice were to
send her full state to him, at a cost of $S(A)$, then he ends up having
entropy $S(AB)$ -- in the quantum world, after you receive negative
information, you know less.

The primitive which (optimally) transfers partial information
we call \emph{quantum state merging}, as Alice's state is effectively merged with
Bob's state, arriving at his site.  With this primitive in hand, one can gain
a systematic understanding of 
quantum network theory, including several important applications
such as distributed compression,  multiple
access channels and assisted entanglement distillation (localizable entanglement),
and compression with quantum side information.  

The purpose of the current paper is to provide full proofs for the result of
\cite{SW-nature}.  In Section \ref{sec:merge} we formally define the notion of
quantum state merging, and state the main result.
In Section \ref{sec:one-shot} we exhibit a general condition to ensure
state merging and derive a one-shot protocol based on random measurements.
In Section \ref{sec:main} we prove the main theorem,
show that our protocol has the optimal classical communication rate, and
provide a heuristic explanation of why the conditional entropy comes into play.

Once the primitive of state merging has been put on a firm footing, we are able to
use it to solve a number of previously intractable problems.  A broad outline of
these applications was given in~\cite{SW-nature}, and here we provide more details.
In  Section \ref{sec:distcomp} we look at the problem of distributed compression,
where several parties at different sites individually compress a source, which is then
decoded by a single party.  It is found that the parties can compress at the ideal
rate of the total entropy, even though they are distributed.  In
Section \ref{sec:sideinfo}, we look at 
noiseless coding with side information, i.e.~we 
consider the problem where one party (Alice) wishes
to compress her state to send to a decoder, and a second party (Bob) who holds part of the
total state can aid her by sending part of his state.  The decoder only wishes
to decode the state of Alice, while Bob's state is only used to help in the
decoding.  As a corollary, we find that if there is a single encoder Alice 
who has access to side information, then this can help her in sending information
to a decoder, a situation impossible in the classical case. 

Next, in Section \ref{sec:eofa}, we treat entanglement of assistance~\cite{entass} in the case
of many helping parties (a concept similar to localizable entanglement~\cite{vpc2004}).
A pure state is shared by many parties, and the goal is to distill the maximum
amount of entanglement between two of the parties.  The other parties can aid
in this distillation through local operations and classical communication.  We find
that state merging gives the optimal rate of distillation.

We then consider the quantum multiple access channel, in Section
\ref{sec:multiply-multiple}. Two parties, Alice and Bob wish
to send quantum states to a decoder through a channel which acts on both their states.
We find optimal rates using state merging, and derive the full rate region.  We are
also able to provide an interpretation to the longstanding puzzle of
negative coherent information in the formula for the capacity of the quantum channel.  
Namely, if one party's rate is negative, than this is the amount of entanglement
he or she must invest in order to help the other party achieve the
maximum rate. 

Before concluding in Section \ref{sec:conclusion}, we provide a quick and intuitive
proof of strong subadditivity using state merging in Section \ref{sec:strongsa}.

\section{State merging: concept, definitions and main result}
\label{sec:merge}
Consider a source emitting a sequence of \emph{unknown} bipartite 
pure states $\ket{\psi_1}_{AB}\ket{\psi_2}_{AB}\ket{\psi_3}_{AB}\ldots$
from a distribution, with average density matrix $\rho_{AB}$.
As with Schumacher compression, we assume the density matrix
of the source is known to the two parties Alice and Bob, but they don't know the ensemble
which realises it. I.e., for any given state they possess, the state is unknown,
although the statistics of the source are.
We are interested in information theoretic quantities, and
in particular, we are interested in quantifying quantum information.
We thus allow free classical
communication between the two parties, and consider many copies $n$ of the
state $\rho_{AB}$.  We now ask how much quantum communication is
needed for Alice to transfer the unknown sequence of states
$\ket{\psi_1}_{AB}\ket{\psi_2}_{AB}\ket{\psi_3}_{AB}\ldots$ to Bob's site. 
This we call {\it quantum state merging}. Notice that because
classical communication is free, we can replace quantum communication
by entanglement due to teleportation~\cite{teleportation} -- this will be
a more convenient way of accounting for the quantum resources.
Faithful state merging
means that the fidelity of the sequence of states is kept
for any realisation of the density matrix.

There is an equivalent, yet more elegant way to conceive of this problem.  We imagine that the
state $\rho_{AB}$ is part of a larger pure state 
$\psi_{ABR} = \proj{\psi}_{ABR}$, with a state vector
$\ket\psi_{ABR}$ which also lives on a reference
(or environment) system $R$.  Faithful state transfer means that the transferred state has high
fidelity with the original state $\ket\psi_{ABR}^{\ot n}$.  More formally,
we define:

\begin{definition}[State merging]
  \label{def:merging}
  Consider a pure state $\ket\Psi_{\tA\tB\tR}$ shared between two parties
  $\tA$, $\tB$ and a reference $\tR$. Let Alice and Bob have further registers
  $A_0,\,A_1$ and $B_0,\,B_1$, respectively. We call a joint operation
  ${\cal M}:\tA A_0 \ot \tB B_0 \longrightarrow A_1 \ot B_1\tB'\tB$
  \emph{state merging of $\Psi$ with error $\epsilon$}, if
  it is LOCC and, with
  $\rho_{A_1 B_1\tB'\tB\tR}
     = ({\cal M}\ot\ID_R) \bigl(\Psi_{\tA\tB\tR}\ot(\Phi_K)_{A_0B_0} \bigr)$,
  \be
    \label{eq:def-merg}
    F\Bigl( \rho_{A_1 B_1\tB'\tB\tR}, (\Phi_L)_{A_1B_1}\ot\Psi_{\tB'\tB\tR} \Bigr)
                                                                  \geq 1-\epsilon,
  \ee
  with maximally entangled states $\Phi_K$, $\Phi_L$ on $A_0B_0$, $A_1B_1$ of
  Schmidt rank $K$, $L$, respectively. Here, $\tB'$ is a local ancilla
  of Bob's of the same size as $\tA$.
  The number $\log K-\log L$ is called the \emph{entanglement cost} of the protocol.
  \par
  In the case of many copies of the same state, $\Psi=\psi^{\ot n}$,
  we call $\frac{1}{n}(\log K-\log L)$ the \emph{entanglement rate}
  of the protocol. A real number $R$ is called an \emph{achievable rate}
  if there exist, for $n\rightarrow\infty$, merging protocols of rate
  approaching $R$ and error approaching $0$. The smallest achievable
  rate is the \emph{merging cost of $\psi$}.
\end{definition}

The main purpose of this paper is to prove in detail the result announced in 
\cite{SW-nature}, namely, that the merging cost
is equal to the conditional entropy of the state $\rho_{AB}$ shared by Alice and Bob,
$S(A|B)=S(B)-S(AB)$.
\begin{theorem}[Quantum State Merging]
  \label{thm:merging}
  For a state $\rho_{AB}$ shared by Alice and Bob, the 
  entanglement cost of merging is equal to the quantum conditional 
  entropy $S(A|B)=S(B)-S(AB)$, in the following sense. When the  $S(A|B)$  is 
  positive, then merging is possible if and only if $R > S(A|B)$ ebits per input copy 
  are provided. When $S(A|B)$ is  negative, then merging is possible 
  by local operations and classical communication,
  and moreover $R< - S(A|B)$ maximally entangled states are obtained per input copy.
\end{theorem}

Our strategy of proof will be the following. We first show that if   
the quantity is negative, then merging can be done by LOCC (indeed,
with only one-way communication from Alice to Bob),
and the entanglement rate that can be obtained is equal to minus the conditional entropy. 
Using this we will show that in the case of positive conditional entropy
it is enough to spend $S(A|B)$ ebits of entanglement.

Finally, we will show that the rates given by the conditional entropy are optimal. 
We will also show that the classical communication cost is equal to the quantum
mutual information between Alice and the reference system $R$,
\be
  I(A:R)=S(A)+S(R)-S(AR)
  \label{eq:mutualae}
\ee
and prove its optimality.

\section{One-shot state merging}
\label{sec:one-shot}
In this section, we first formulate a general sufficient condition
on a measurement of Alice that ensures that Bob can complete state merging
by local operations; then we show how random measurements succeed
with high probability in realising this condition.

\subsection{Condition for merging with one-way LOCC}
\label{ss:merging-conditions}
Here we will provide a condition that is sufficient to 
obtain state merging with only LOCC. We formulate it in the
one-shot setting of definition~\ref{def:merging}.
It is based on Alice performing a measurement which takes the original state
$\Psi_{\tA\tB\tR}$ to another pure state, with the essential features that:
(1) the reference system $\tR$ is unchanged, and (2) Alice's and the Reference's states
are in product form.  Since all purifications are equal up to a local unitaries,
this implies that Bob can perform a local unitary which transforms his state
into $\rho_{\tA\tB}$.

More formally, we consider a protocol, whose basic constituent is Alice's 
incomplete measurement given by Kraus operators $P_j$ mapping
$\tA$ to $A_1$ (in our actual solution,
it will be a von Neumann measurement followed by a unitary). 
Given the outcome was $j$, the state $\Psi_{\tA\tB\tR}$  
collapses to a state which we will denote by $\Psi^j_{A_1\tB\tR}$,
\be
  |\Psi^j\>_{A_1\tB\tR} = \frac{1}{\sqrt{p_j}} (P_j\ot I_{\tB\tR}) |\Psi\>_{\tA\tB\tR},
\ee
where $p_j$ is probability of obtaining outcome $j$,
\be
p_j = \<\Psi| (P_j^\dagger P_j \ot  I_{\tB\tR}) |\Psi\>.
\ee
Suppose for the moment that $|\Psi^j\>_{A_1\tB\tR}$ has the property
\be
  \rho^j_{A_1\tR} = \tau_{A_1} \ot \rho_{\tR},
  \label{eq:product}
\ee
where $\rho^j_{A_1\tR}$ is the reduced density matrix of $\Psi^j_{A_1\tB\tR}$, 
$\tau_{A_1}$ is the maximally mixed state of dimension $L$ on Alice's system $A_1$,
and $\rho_{\tR}$ is the reduced density matrix of the original state $\Psi_{\tA\tB\tR}$. 
Then (see~\cite{SchumacherW01-approx}) there exists an isometry
$U_j:\tB \longrightarrow B_1\tB'\tB$ on Bob's side, such that
\be
  (I_{A_1\tR} \ot U_j) \ket{\Psi^j}_{A_1\tB\tR} 
                           = \ket{\Phi_L}_{A_1B_1} \ot \ket{\Psi}_{\tB'\tB\tR},
\ee
where $\ket{\Psi}_{\tB'\tB\tR}$ is the original state $\ket{\Psi}_{\tA\tB\tR}$
with the system $\tB'$ substituted for $\tA$.
This is because $\Psi$ is the purification of $\rho_{\tR}$ and
$\Phi_L$ that of $\tau_{A_1}$, so both $\Psi^j_{A_1\tB\tR}$ and
$(\Phi_L)_{A_1B_1} \ot \Psi_{\tB'\tB\tR}$ are purifications of
$\tau_{A_1} \ot \rho_{\tR}$. Hence, by Uhlmann's theorem, they
are related by a unitary on Bob's system.
%

Since we require fidelity approaching $1$ only in the asymptotic limit, 
we obtain the following merging condition:
\begin{proposition}[Merging condition]
\label{prop:merging-cond1}
Consider Alice's measurement with outcomes $j$, which occur with
probability $p_j$. Denote the state after the measurement result $j$ 
was obtained by $\ket{\Psi^j}_{A_1\tB\tR}$,
and its reduced density matrix by $\rho^j_{A_1\tR}$.
The following condition implies the existence of a
merging protocol with entanglement cost $-\log L$ and
error $2\sqrt{\ep}$: that the so-called \emph{quantum error}
$Q_e$ satisfies
\be
  Q_e := \sum_j p_j \bigl\| \rho^j_{A_1\tR} - \tau_{A_1} \ot \rho_{\tR} \bigr\|_1 \leq \ep,
  \label{eq:merging-condition1}
\ee
where $\tau_{A_1}$ is the maximally mixed state of dimension $L$ on $A_1$. 
\end{proposition}

\begin{beweis}
The proof is based on the above considerations concerning the 
ideal situation. Using the relation eq.~(\ref{eq:fuchs}) between the
trace distance and the fidelity, we get
\be
  \sum_j p_j \sqrt{F\bigl( \rho^j_{A_1\tR},\tau_{A_1} \ot \rho_{\tR} \bigr)}
                                                                 \geq 1-\frac{\ep}{2}.
\ee
Then, by Uhlmann's theorem~\cite{uhlmann-fid,jozsa-fid}
there exist isometries $U_j$ of Bob such that
\be
  F\bigl( \rho^j_{A_1\tR},\tau_{A_1} \ot \rho_{\tR} \bigr)
           = F\bigl( (I_{A_1\tR} \ot U_j) \ket{\Psi^j}_{A_1\tB\tR},
                       \ket{\Phi_L}_{A_1B_1} \ot \ket{\Psi}_{\tB'\tB\tR} \bigr),
\ee
hence
\begin{equation}\begin{split}
  \sum_j &p_j F\bigl( (I_{A_1\tR} \ot U_j) \ket{\Psi^j}_{A_1\tB\tR},
                       \ket{\Phi_L}_{A_1B_1} \ot \ket{\Psi}_{\tB'\tB\tR} \bigr)             \\
         &\phantom{===}
          \geq \left( \sum_j p_j \sqrt{F\bigl( (I_{A_1\tR} \ot U_j)\ket{\Psi^j}_{A_1\tB\tR},
                       \ket{\Phi_L}_{A_1B_1} \ot \ket{\Psi}_{\tB'\tB\tR} \bigr)} \right)^2  \\
         &\phantom{===}
          \geq \left(1-\frac{\ep}{2}\right)^2 \geq 1-\ep.
\end{split}\end{equation}
So, with the output state of the protocol,
\be
  \rho_{A_1 B_1\tA'\tB\tR} = \sum_j (I_{A_1\tR} \ot U_j)
                                       \proj{\Psi^j}_{A_1\tB\tR}
                                    (I_{A_1\tR} \ot U_j)^\dagger,
\ee
we obtain
\be
  F\bigl( \rho_{A_1 B_1\tA'\tB\tR},(\Phi_L)_{A_1B_1} \ot \Psi_{\tB'\tB\tR} \bigr)
                                                                        \geq 1-\ep.
\ee
And using the relation (\ref{eq:fuchs}) between fidelity and trace
distance once more, we arrive at
\be
  \bigl\| \rho_{A_1 B_1\tA'\tB\tR} - (\Phi_L)_{A_1B_1} \ot \Psi_{\tB'\tB\tR} \bigr\|_1
                                                                   \leq 2\sqrt{\epsilon},
\ee
which concludes the proof.
\end{beweis}

Note that for any protocol which achieves merging, the
condition (\ref{eq:merging-condition1})
must necessarily be met at some stage of the protocol.  
This is because in order for the final state
to be close to the original state, $\rho_{\tR}$ must 
necessarily be virtually unchanged, and in order
for the state to be at Bob's site, Alice's state must 
necessarily be in a product state with the
reference system $\tR$.

\subsection{One-shot merging by random measurement}
\label{ss:one-shot}
Here we will prove an abstract, one-shot version of the main theorem,
showing that a random orthogonal measurement of rank-$L$ projectors
(and a little remainder) achieves merging.
\begin{proposition}[One-shot merging]
  \label{prop:one-shot:merge}
  Let $\Psi_{\tA\tB\tR}$ be a pure state, with local dimensions
  $d_{\tA}$, $d_{\tB}$, $d_{\tR}$, and $\tr\rho_{\tB}^2 \leq \frac{1}{D}$.
  Then there exists a POVM consisting of $N = \left\lfloor \frac{d_{\tA}}{L} \right\rfloor$
  projectors of rank $L$ and one of rank $L' = d_{\tA}-NL < L$ such that
  \be
    \label{eq:Qe-ave}
    Q_e \leq 2\sqrt{L\frac{d_{\tR}}{D}} + 2\frac{L}{d_{\tA}},
  \ee
  and there is a merging protocol with error at most
  $2\sqrt{2\sqrt{L\frac{d_{\tR}}{D}} + 2\frac{L}{d_{\tA}}}$.

  In fact, by choosing the measurement at random according to the
  Haar measure on $\tA$, the expectation of the left hand
  side of eq.~(\ref{eq:Qe-ave}) is upper bounded by the right hand side.
\end{proposition}

\begin{remark}
  Let us explain here briefly how we will use the Lemma in the proof
  of Theorem \ref{thm:merging} in the case of negative $S(A|B)$. 
  Namely, we will apply this Lemma with the following parameters: 
  $d_{\tR}\approx 2^{nS_R}=2^{nS_{AB}}$, 
  $d_{\tA}\approx 2^{nS_R}=2^{nS(A)}$, $D\approx 2^{nS(R)}=2^{nS(B)}$, 
  where $n$ is the number of copies of 
  initial state $\psi_{ABR}$  shared by Alice, Bob and reference system.
  Moreover $L$ will be related to the rate $ r $ of singlets obtained
  between Alice and Bob in the process of merging: 
  $L\approx 2^{n r}$.  Then the expression for quantum error will be 
  \be
    Q_e \approx 2^{\frac{1}{2} n (S(AB) - S(B) + r)} + 2^{n(r - S(A))+1}
  \ee
  Thus if only $r<S(AB) - S(B)$, then the quantum error will decay
  exponentially with  $n$.
\end{remark}

The crucial technical result in the proof of
Proposition~\ref{prop:one-shot:merge} will be the following statement
about random (Haar distributed) rank-$L$ projectors:
\begin{lemma}
  \label{lemma:crucial-average}
  Let $P:\tA\longrightarrow A_1$ be a random partial
  isometry of rank $L$, i.e.~$P^\dagger P$ is a projection onto
  a $L$-dimensional subspace of $\tA$. For example, one might
  put $P=P_0 U$ with some fixed rank $L$-projector $P_0$
  onto a subspace $A_1$ of $\tA$, and a Haar distributed
  unitary $U$ on $\tA$.
  For the subnormalized density matrix
  $$\omega_{A_1\tR} = (P\ot I_{\tR})\rho_{\tA\tR}(P\ot I_{\tR})^\dagger,$$
  observe that its average over unitaries $U$ is
  $$\langle \omega_{A_1\tR} \rangle = \frac{L}{d_{\tA}}\tau_{A_1}\ot\rho_{\tR}.$$
  And we have:
  \begin{align}
    \label{eq:2-average}
    \left\langle \left\| \omega_{A_1\tR}
                        - \frac{L}{d_{\tA}}\tau_{A_1}\ot\rho_{\tR} \right\|_2^2 \right\rangle
                                   &\leq \frac{L^2}{d_{\tA}^2}\frac{1}{D}, \\
    \label{eq:1-average}
    \left\langle \left\| \omega_{A_1\tR}
                        - \frac{L}{d_{\tA}}\tau_{A_1}\ot\rho_{\tR} \right\|_1 \right\rangle
                                   &\leq \frac{L}{d_{\tA}}\sqrt{L\frac{d_{\tR}}{D}}.
  \end{align}
\end{lemma}
\begin{beweis}
  In Appendix~\ref{app:misc} we recall the basic properties of the trace norm $\|\cdot\|_1$
  and the Hilbert-Schmidt norm $\|\cdot\|_2$.
  From there (Lemma~\ref{lem:norms})
  we take that $\|X\|_1 \leq \sqrt{d}\|X\|_2$ for
  an operator on a $d$-dimensional space. This, and the concavity
  of the square root function, show that eq.~(\ref{eq:2-average})
  implies eq.~(\ref{eq:1-average}).

  To prove eq.~(\ref{eq:2-average}), we use the fact that it has the
  form of a variance, so
  \begin{equation}\begin{split}
    \left\langle \left\| \omega_{A_1\tR}
                        - \frac{L}{d_{\tA}}\tau_{A_1}\ot\rho_{\tR} \right\|_2^2 \right\rangle
                &= \left\langle \left\| \omega_{A_1\tR}
                        - \langle \omega_{A_1\tR} \rangle \right\|_2^2 \right\rangle \\
                &= \bigl\langle \tr\omega_{A_1\tR}^2 \bigr\rangle
                        - \tr{\langle\omega_{A_1\tR}\rangle}^2 \\
                &= \bigl\langle \tr\omega_{A_1\tR}^2 \bigr\rangle
                        - \frac{L^2}{d_{\tA}^2}\frac{1}{L}\tr\rho_{\tR}^2.
                                                      \label{eq:steppingstone}
  \end{split}\end{equation}

  To evaluate the average of $\tr\omega_{A_1\tR}^2$, we use the 
  well-known equation
  \be
    \tr\omega_{A_1\tR}^2 = \tr\bigl( (\omega_{A_1\tR}\ot\omega_{A_1\tR})
                                              (F_{A_1A_1}\ot F_{\tR\tR}) \bigr),
  \ee
  where we have introduced copies of all systems involved, and with the
  swap (or flip) operator $F$ exchanging the two systems.
  (Note that $F_{\tA\tR,\tA\tR} = F_{\tA\tA} \ot F_{\tR\tR}$.)
  With this, and w.l.o.g.~assuming that $A_1$ is a subspace of $\tA$,
  \begin{equation}\begin{split}
    \bigl\langle \tr\omega_{A_1\tR}^2 \bigr\rangle
           &= \Bigl\langle \tr\bigl( (\omega_{A_1\tR}\ot\omega_{A_1\tR})
                                              (F_{A_1A_1}\ot F_{\tR\tR}) \bigr) \Bigr\rangle
                                                                                   \\
           &= \Bigl\langle \tr\bigl( ({UU}_{\tA\tA}\ot I_{\tR\tR})
                                              (\rho_{\tA\tR}\ot\rho_{\tA\tR})
                                     ({UU}_{\tA\tA}\ot I_{\tR\tR})^\dagger
                                              (F_{A_1A_1}\ot F_{\tR\tR}) \bigr) \Bigr\rangle
                                                                                   \\
           &= \tr\Bigl( (\rho_{\tA\tR}\ot\rho_{\tA\tR})
                        \bigl\langle ({UU}_{\tA\tA}\ot I_{\tR\tR})^\dagger
                                              (F_{A_1A_1}\ot F_{\tR\tR}) 
                                     ({UU}_{\tA\tA}\ot I_{\tR\tR}) \bigr\rangle \Bigr)      
                                                                                   \\
           &= \tr\Bigl( (\rho_{\tA\tR}\ot\rho_{\tA\tR})
                        \bigl( \bigl\langle ({UU}_{\tA\tA})^\dagger 
                                            F_{A_1A_1} ({UU}_{\tA\tA}) \bigr\rangle
                        \ot F_{\tR\tR} \bigr) \Bigr),                        \label{eq:phew}
  \end{split}\end{equation}
  where we have used the shorthand ${UU}_{\tA\tA} := U_{\tA}\ot U_{\tA}$.
  In Appendix \ref{app:twirl} we demonstrate, how using elementary arguments
  from the representation theory of $U\ot U$, one can calculate that
  \be
    \label{eq:twirled}
    \bigl\langle ({UU}_{\tA\tA})^\dagger F_{A_1A_1} ({UU}_{\tA\tA}) \bigr\rangle
      = \frac{L}{d_{\tA}}\frac{d_{\tA}-L}{d_{\tA}^2-1}I_{\tA\tA}
       + \frac{L}{d_{\tA}}\frac{Ld_{\tA}-1}{d_{\tA}^2-1}F_{\tA\tA}.
  \ee
  Inserting this into eq.~(\ref{eq:phew}) gives
  \begin{equation}\begin{split}
    \bigl\langle \tr\omega_{A_1\tR}^2 \bigr\rangle 
            &=    \frac{L}{d_{\tA}}\frac{d_{\tA}-L}{d_{\tA}^2-1}\tr\rho_{\tR}^2
                  + \frac{L}{d_{\tA}}\frac{Ld_{\tA}-1}{d_{\tA}^2-1}\tr\rho_{\tA\tR}^2 \\
            &\leq \frac{L}{d_{\tA}^2}\tr\rho_{\tR}^2
                  + \frac{L^2}{d_{\tA}^2}\frac{1}{D},
  \end{split}\end{equation}
  and looking at eq.~(\ref{eq:steppingstone}) we are done.
\end{beweis}

\begin{beweis}[of Proposition~\ref{prop:one-shot:merge}]
  Fix a random measurement according to the description of the
  Proposition. One way of doing this is picking $N$ fixed
  orthogonal subspaces of dimension $L$, and one of dimension
  $L' = d_{\tA}-NL < L$. The projectors onto these subspaces
  followed by a fixed unitary mapping it to $A_1$ we denote
  by $Q_j$, $j=0,\ldots,N$. Then put $P_j := Q_jU$ with
  a Haar distributed random unitary $U$ on $\tA$.

  Then, by Lemma~\ref{lemma:crucial-average}, with
  $\omega_{A_1\tR}^j = (P_j\ot I_{\tR})\rho_{\tA\tR}(P_j\ot I_{\tR})^\dagger$,
  \begin{equation}
    \label{eq:penultimate}
      \left\langle \sum_{j=1}^N \left\| \omega_{A_1\tR}^j
                           - \frac{L}{d_{\tA}}\tau_{A_1}\ot\rho_{\tR} \right\|_1 \right\rangle
      \leq N\frac{L}{d_{\tA}}\sqrt{L\frac{d_{\tR}}{D}}
      \leq \sqrt{L\frac{d_{\tR}}{D}}.
  \end{equation}
  This is almost what we want, except that we haven't taken into account
  the normalisation: with $p_j=\tr\omega_{A_1\tR}^j$ and
  $\rho_{A_1\tR}^j = \frac{1}{p_j}\omega_{A_1\tR}^j$,
  we need to argue that on average, the $p_j$ are close to $\frac{L}{d_{\tA}}$.
  Indeed, eq.~(\ref{eq:penultimate}) implies
  \be
    \left\langle \sum_{j=1}^N \left| p_j - \frac{L}{d_{\tA}} \right| \right\rangle
                                                       \leq \sqrt{L\frac{d_{\tR}}{D}},
  \ee
  hence we obtain
  \be
      \left\langle \sum_{j=1}^N p_j \left\| \rho_{A_1\tR}^j
                           - \tau_{A_1}\ot\rho_{\tR} \right\|_1 \right\rangle
                                                  \leq 2\sqrt{L\frac{d_{\tR}}{D}}.
  \ee
  Finally, it is clear that
  $\bigl\langle \tr(\rho_{\tA\tR}P_0) \bigr\rangle
   = \frac{L'}{d_{\tA}} < \frac{L}{d_{\tA}}$,
  and since the trace distance of two states is at most $2$,
  we get the result as advertised, because the quantum error is
  composed of the probability of hitting $P_0$ and the
  sum of the error terms of the $P_j$, weighted by their probabilities.
  Now we can apply Proposition~\ref{prop:merging-cond1}.
\end{beweis}

So, if $d_{\tR} \ll D$ there is a merging LOCC protocol with
small error and entanglement cost up to $\log d_{\tR} - \log D$
(i.e., the negative of this is the amount of entanglement produced).
If $d_{\tR} \not\ll D$, consider the state $\Psi_{\tA\tB\tR}\ot(\Phi_K)_{A_0B_0}$
with a maximally entangled state of Schmidt rank $K \gg d_{\tR}/D$.
Now merging is possible (with $L=1$); the entanglement cost is $\log K$,
and it can be made as small as $\log d_{\tR} - \log D$.

\section{Proof of the main theorem}

\label{sec:main}

\subsection{Achievability of merging}

\begin{beweis}[of Theorem~\ref{thm:merging}]
We will first prove the direct part saying that the rates are achievable.
Consider $n$ copies of the state $\ket{\psi}_{ABR}$, and assume
first that $S(A|B) < 0$.

We would like to use our one-shot version,
Proposition~\ref{prop:one-shot:merge}, but cannot do so directly,
since the dimension $d_R^n$ and the number $(\tr\rho_R^2)^n$
are not information theoretically meaningful.

Instead, we consider the vector $\ket{\Omega}_{\tA\tB\tR}$ and state
$\ket{\Psi}_{\tA\tB\tR}$, with
\be
  \ket{\Omega}_{\tA\tB\tR} := (\Pi_{\tA}\ot\Pi_{\tB}\ot\Pi_{\tR})\ket{\psi}_{ABR}^{\ot n},
  \qquad
  \ket{\Psi}_{\tA\tB\tR} := \frac{1}{\bra{\Omega}\Omega\rangle}\ket{\Omega}_{\tA\tB\tR},
\ee
where $\tA$, $\tB$ and $\tR$ are the typical subspaces of $A^n$,
$B^n$ and $R^n$, respectively, and $\Pi_{\tA}$, etc. are the projection
operators onto these typical subspaces.
In Appendix~\ref{app:typicality} we explain what is necessary to know
about typicality, in particular we have:
\be
  \label{eq:normalisation}
  \bra{\Omega}\Omega\rangle
   = \bra{\psi}^{\ot n} (\Pi_{\tA}\ot\Pi_{\tB}\ot\Pi_{\tR}) \ket{\psi}^{\ot n}
                                                                     \geq 1-\ep,
\ee
for any $\ep>0$ and large enough $n$.
Indeed, we can choose $\ep = 3\exp(-c\delta^2 n)$ with some constant $c$,
where $\delta>0$ is a typicality parameter; namely from
eq.~(\ref{eq:C1}) in Appendix \ref{app:typicality} we have
$\tr\rho_A^{\ot n}\Pi_{\tA},\,
 \tr\rho_B^{\ot n}\Pi_{\tB},\,
 \tr\rho_R^{\ot n}\Pi_{\tR}\, \geq 1-\exp(-c\delta^2 n)$.
We obtain the bound (\ref{eq:normalisation}) from observing
\be
  I_{A^n} \ot I_{B^n} \ot I_{R^n} - \Pi_{\tA} \ot \Pi_{\tB} \ot \Pi_{\tR}
     \leq (I_{A^n}-\Pi_{\tA}) \ot (I_{B^n}-\Pi_{\tB}) \ot (I_{R^n}-\Pi_{\tR}).
\ee
Furthermore, with $\Omega = \proj{\Omega}$, we have (using
eqs.~(\ref{eq:C6}), (\ref{eq:C5}) and (\ref{eq:C2})
in Appendix \ref{app:typicality})
\begin{align}
  \rk \Omega_{\tA} &\geq (1-\ep)2^{n[S(A)-\delta]},     \nonumber \\
  \rk \Omega_{\tR} &\leq 2^{n[S(R)+\delta]},                      \\
  \Omega_{\tB}     &\leq \Pi_{\tB}\rho_B^{\ot n}\Pi_{\tB}
                    \leq 2^{-n[S(B)-\delta]} \Pi_{\tB}. \nonumber
\end{align}
Hence we get, for the normalized $\Psi_{\tA\tB\tR}$,
\be
  d_{\tA} \geq (1-\ep)2^{n[S(A)-\delta]}, \quad
  d_{\tR} \leq 2^{n[S(R)+\delta]},        \quad
  D       \geq (1-\ep)^2 2^{n[S(B)-\delta]}.
\ee
By the gentle measurement Lemma~\ref{lemma:gentle} (see Appendix~\ref{app:misc}),
we obtain from eq.~(\ref{eq:normalisation})
\be
  \label{eq:approximation}
  \bigl\| \psi_{ABR}^{\ot n} - \Omega_{\tA\tB\tR} \bigr\|_1 \leq 2\sqrt{\ep},
  \text{ hence }
  \bigl\| \psi_{ABR}^{\ot n} - \Psi_{\tA\tB\tR} \bigr\|_1 \leq 4\sqrt{\ep}.
\ee
Now Alice and Bob follow a merging protocol as if they had $\Psi_{\tA\tB\tR}$,
and with $L=2^{n[S(B)-S(R)-3\delta]}$. If the state were actually
$\Psi_{\tA\tB\tR}$, the quantum error would be
\be
  Q_e \leq 2\sqrt{L\frac{d_{\tR}}{D}} + 2\frac{L}{d_{\tA}}
      \leq \frac{2}{1-\ep} 2^{-n\delta/2} + 2^{1-2n\delta}.
\ee
(Observe that $S(B)-S(R) \leq S(A)$ by subadditivity.)
So, by Proposition~\ref{prop:merging-cond1} we would get
a merging protocol with error $O(2^{-n\delta/4})$.
By eq.~(\ref{eq:approximation}), running the same protocol on
$\psi_{ABR}^{\ot n}$, we obtain an error of
$O(2^{-n\delta/4}) + O(2^{-cn\delta^2/2})$,
which vanishes exponentially as $n\rightarrow\infty$.
Since $\delta>0$ was arbitrary, the direct part follows.

It remains to consider the case when $S(A|B)$ is non-negative.
Here, Alice and Bob share additionally $n(S(A|B)+\Delta)$ 
maximally entangled states. Each ebit
contributes conditional entropy $-1$,
so that the final state has negative conditional entropy $-n\Delta$. 
Then however merging can be done by LOCC, as we have proven above. 

\begin{remark}
  \label{rem:whatever}
  Note that despite the generality of the definition of merging,
  our protocol is much more special. The definition allows
  to start end end with certain amounts of ebits, but the amount
  charged is only the difference, so that it would be conceivable
  that to achieve the conditional entropy some catalytic
  use of entanglement is necessary.
  However, our protocol either needs no initial entanglement
  and outputs some (if $S(A|B)<0$) or produces no entanglement
  but needs some initially (if $S(A|B)\geq 0$).
\end{remark}

\subsection{Merging is optimal}

Let us now turn to the converse part. 
The essence of the proof is that entanglement cannot increase under local operations and 
classical communication and transmission of $n$ qubits more than by $n$ \cite{LoPopescu}. 
We will consider preservation of Bob's entanglement 
with Alice and the Reference.
The initial entanglement $E_{in}$ includes the entanglement of the
shared state plus any initial resource
of pure entanglement $\log K$. Initially, it is
$n S(B) + \log K$ as the initial state was just $\psi_{ABR}^{\ot n}$.
%
The final entanglement $E_{out}$ includes the entanglement of the
final state plus the final resource, 
$\log L$ bits of pure state entanglement, and is
\be
E_{out}\approx nS(AB) + n \log L .
\ee
Since Alice and Bob used only LOCC 
operations, we have 
\be
E_{out}\leq E_{in}
\ee
as entanglement could only decrease, giving $R=\log K-\log L < S(AB)-S(B)$. 

In more detail, assume 
$L\leq 2^{O(n)}$ for technical reasons. The LOCC protocol (which is also LOCC between
Bob and Alice$+$Reference) can be thought of as generating an
ensemble $\{ \varphi^k_{A_1B_1{B'}^nB^nR^n} , q_k \}$ of pure states.
Monotonicity of the entropy of entanglement under LOCC
\cite{BBPS1996} means
\be
  \label{eq:monotonous}
  n S(B) + \log K \geq \sum_k q_k S\bigl( \varphi^k_{B_1{B'}^nB^n} \bigr).
\ee
The condition (\ref{eq:def-merg}) for successful merging translates into
\be
  \sum_k q_k F\bigl( \varphi^k_{A_1B_1{B'}^nB^nR^n},
                      (\Phi_L)_{A_1B_1}\ot \psi_{B'BR}^{\ot n} \bigr) \geq 1-\ep,
\ee
thanks to the linearity of the fidelity when one argument is pure.
Using eq.~(\ref{eq:fuchs}) in Appendix \ref{app:misc} this yields
\be
  \sum_k q_k \bigl\| \varphi^k_{A_1B_1{B'}^nB^nR^n}
                      - (\Phi_L)_{A_1B_1}\ot \psi_{B'BR}^{\ot n} \bigr\|_1 \leq 2\sqrt{\ep},
\ee
hence by monotonicity of the trace norm under partial tracing,
\be
  \label{eq:what-is-that}
  \sum_k q_k \bigl\| \varphi^k_{B_1{B'}^nB^n}
                      - \tau_{A_1} \ot \rho_{B'B}^{\ot n} \bigr\|_1 \leq 2\sqrt{\ep}.
\ee
By Fannes' inequality (stated as Lemma~\ref{lem:fannes} in
Appendix \ref{app:misc}), this finally gives
\be
  \sum_k q_k \bigl| S(\varphi^k_{B_1{B'}^nB^n}) - \log L - nS(AB) \bigr| 
                   \leq \bigl( \log L+n\log d_A+n\log d_B \bigr) \eta(2\sqrt{\ep})
                   \leq O(n) \eta(2\sqrt{\ep}),
\ee
using the concavity of the $\eta$-function. With eq.~(\ref{eq:monotonous}),
we thus get
\be
  \frac{1}{n}(\log K - \log L) \geq S(A|B) - O(1)\eta(2\sqrt{\ep}),
\ee
which results in the converse when $n\rightarrow\infty$ and
$\ep\rightarrow 0$.
\end{beweis}

\subsection{Classical communication cost of merging}
\label{ss:cc-merg}
In our protocol for quantum state merging, the amount of classical communication that
Alice needs to send Bob is given by the number of possible measurement outcomes:
at most $\frac{d_{\tA}d_{\tR}}{D}+1$, which in the i.i.d.~case $\psi_{ABR}^{\ot n}$
means a rate of $S(A)+S(R)-S(B) = I(A:R)$.
Note that this is true regardless of $S(A|B) \geq 0$ or $S(A|B) < 0$.

We now show that this amount
of communication is needed, and thus our protocol is communication optimal.
\begin{theorem}
\label{thm:mergingcc}
For a state $\ket{\psi}_{ABR}$ shared by Alice, Bob and the Reference, the 
classical communication cost of merging is equal to the quantum mutual information 
between Alice and the reference system $R$, $I(A:R)=S(A)+S(R)-S(AR)$.
\end{theorem}
\begin{beweis}
We will first need to take a short digression.  
Consider a protocol which achieves merging with a entanglement rate
$R_q$ and classical communication at rate $R_c$.  Now let us imagine that
the parties do not have access to a classical channel, so must send all their
classical communication via the quantum channel, encoded into qubits.
This gives a fully quantum version of merging~\cite{igor-personal} similar to
the ``mother protocol'' (see~\cite{igor-fqsw} for an alternative,
direct proof). If $R_q=S(A|B)$ and $R_c=I(A:R)$,
we have, in the ``sloppy'' notation of~\cite{famtrade},
\beq
  \frac{1}{2}I(A:R)[q\rightarrow q] \geq \frac{1}{2}I(A:B)[qq] + \merging,
  \label{eq:mother}
\eeq
where the equation means that a rate of $\frac{1}{2} I(A:R)$ uses of a noiseless
qubit channel $[q\rightarrow q]$,
and it produces $\frac{1}{2} I(A:B)$ bits of shared entanglement
$[qq]$ in addition to achieving state merging from
Alice to Bob. The latter is represented
by $\merging$, i.e.~a identity channel from Alice to Bob working
on the source $\rho_{AB}$.

We briefly sketch how state merging gives the
protocol of eq.~(\ref{eq:mother}). Our merging protocol is expressed in the
resource inequality formalism as
\beq
  S(A|B) [qq] + I(A:R) [c\rightarrow c] \geq \merging,
  \label{eq:merge:RI}
\eeq
where $[c\rightarrow c]$ stands for the communication resource of $1$
classical bit.
Recall that for any state merging protocol, 
the classical communication must be completely decoupled from the sent
state for $|\psi\>_{ABR}$ to remain pure, and thus
it can be recycled as $R_c$ bits of entanglement;
the entanglement can further be used to send quantum states.
This is what the authors of~\cite{dhw-family,famtrade} call {\it Rule I},
where each bit of classical communication (denoted as $[c\rightarrow c]$)
can be made coherent: we denote a coherent classical~\cite{Harrow-ccc}
bit by $[q\rightarrow qq]$.
At the left hand side of an inequality like (\ref{eq:merge:RI}),
{\it Rule I} says that it can be replaced by half a bit
of a quantum channel on the left
and half a bit of shared entanglement on the right hand side
(denoted $\frac{1}{2}[q\rightarrow q] - \frac{1}{2}[qq]$).
One sees this by sending the classical communication used in teleportation as
coherent qubits which are then recycled into entanglement. Thus,
\beq
  [q\rightarrow qq] = \frac{1}{2}[q\rightarrow q]-\frac{1}{2}[qq] .
  \label{eq:rulei}
\eeq
Applying {\it Rule I} of eq.~(\ref{eq:rulei}) to
eq.~(\ref{eq:merge:RI}), and rearranging the terms
gives the mother protocol in the formulation of eq.~(\ref{eq:mother}).

We now show that the mother is an optimal protocol to achieve state merging
in the case when one doesn't have access to a classical channel
(see also~\cite{igor-fqsw}).
We use the fact that a necessary condition for
any state merging protocol is that Alice must completely decouple
herself from the state $|\psi\>_{ABR}$.  This is because the state
needs to be shared by $R$ and $B$ by definition of state merging.

Whatever Alice does, including measurements and processing, 
we may consider coherently, as an operation which takes $\rho_A$ and
some ancillas, and produces a part which gets sent down the quantum channel,
and a part $\rho_{A'}$ she retains.
This results in a state $|\psi'\>_{BB'R}$ which has high fidelity with
$|\psi\>_{ABR}$, plus some entanglement between Alice and Bob.
Now, using standard quantum cryptographic reasoning
originating in~\cite{Ekert91}, if $|\psi'\>_{BB'R}$ is (almost) pure, then
the system $A'$ must be virtually in a product state with $B'BR$.
In particular, the mutual information between the state $\rho_A'$ and
the reference system $R$ must be close to zero. 
Each qubit sent can reduce Alice's mutual information with
the reference system by at most $2$, thus at a minimum, Alice must send
$\frac{1}{2} I(A:R)$ qubits down the quantum channel. This gives the optimality
of Alice's use of the quantum channel in protocol (\ref{eq:mother}).

That at most $\frac{1}{2} I(A:B)$ bits of entanglement are obtainable from the shared
state, when sending $\frac{1}{2} I(A:R)$ qubits, can be easily seen
as follows. Observe that the $\frac{1}{2} I(A:R)[q\rightarrow q]$ on the
left hand side of eq.~(\ref{eq:mother}) can be replaced by
$\frac{1}{2}I(A:R) [qq] + I(A:R) [c\rightarrow c]$ due to teleportation.
If the entanglement rate on the right were larger than
$\frac{1}{2} I(A:B)$, we could perform state merging
with entanglement rate strictly smaller than
$\frac{1}{2}I(A:R) - \frac{1}{2}I(A:B) = S(A|B)$, contradicting the
converse of Theorem~\ref{thm:merging}.

Now, to prove optimality of the classical communication in
eq.~(\ref{eq:merge:RI}), consider a hypothetical state merging protocol
\beq
  R_q [qq]+R_c [c\rightarrow c] \geq \merging 
  \label{eq:genpremom}
\eeq
which we may transform using {\it Rule I}~\cite{dhw-family,famtrade} into
\beq
  \left( R_q-\frac{1}{2}R_c \right) [qq] + \frac{1}{2}R_c [q\rightarrow q] \geq \merging.
\eeq
Comparing this with the mother protocol (\ref{eq:mother}),
we have that $R_c\geq I(A:R)$ by virtue of the optimality of (\ref{eq:mother});
$R_q \geq S(A|B)$ comes out again, as it should.
\end{beweis}

Thus in addition to giving an operational interpretation for the quantum condition entropy, 
merging gives an operational interpretation for the quantum mutual information.
Secondly, the measurement of Alice makes her state completely
product with $R$, thus reinforcing the interpretation 
of quantum mutual information 
as the minimum entropy production of any local
decorrelating process~\cite{GroismanPW04,huge}.
This same quantity is also equal to the amount of
irreversibility of a cyclic process: Bob initially has a state, 
then gives Alice her share
(communicating $S(A)$ qubits),
which is finally merged back to him (communicating $\sacb$ qubits). 
The total quantum communication of this cycle is $I(A:R)$ quantum bits.


\bigskip
Having concluded our proofs regarding state merging, we now turn to its applications.

\section{Distributed compression}
\label{sec:distcomp}
In usual Schumacher compression, a single party Alice, receives a state from a source,
and must compress the states so that they can be faithfully decoded by another party.
For a source  emitting states with density matrix 
$\ra$, this can be done at a rate given by the entropy $\sa$ of the source
\cite{Schumacher1995}.
One can imagine the situation where the states are distributed over many parties,
and have to be compressed individually.  Each party then sends their
compressed share to a decoder who must be able to decode the full state.  
In the classical case, this problem was solved by
Slepian and Wolf~\cite{slepian-wolf} who found that the
total rate for distributed compression could equal the
compression rate when the parties are not distributed. 
In the quantum case, previous results~\cite{Winter1999PhD,adhw2004}
were interpreted as indications that one cannot 
compress at the same rate in the distributed
vs. non-distributed case. However, using state merging,
we will show that formally the same achievable rate region
as in the Slepian-Wolf theorem is obtained

In detail, we assume that the source emits states with average density matrix 
$\r_{A_1A_2...A_m}$, and distributes it over $m$ parties.
The parties wish to compress their shares as much as possible 
so that the full state can be reconstructed by a 
single decoder. We allow
classical side information for free (we will only need classical communication
from each encoder to the joint decoder), and only ask about the rate
$R_i$ of entanglement between the $i^{\text{th}}$ encoder and the decoder.
A tuple $(R_1,\ldots,R_m)$ is \emph{achievable} if there exists an
$(m+1)$-party LOCC procedure taking in the source
$\rho_{A_1\ldots A_m}^{\otimes n}$, purified to a state
$\psi_{RA_1\ldots A_m}^{\otimes n}$, and $n(R+\ep)$ ebits
between $A_i$ and the decoder $B$, such that the final state is
$\rho_{R^nB_1'\ldots B_m'}$ with
\be
  F\bigl( \rho,\psi^{\otimes n} \bigr) \geq 1-\ep,
\ee
and $\ep\rightarrow 0$ as $n\rightarrow \infty$. As always, the reference is passive,
and plays no role in the protocol.  Note that
the rates $R_i$ can be negative here, just as in state merging,
meaning that $n(-R_i+\ep)$ ebits are returned by the protocol.

\begin{figure}[ht]
  \centering
  \includegraphics[width=12cm]{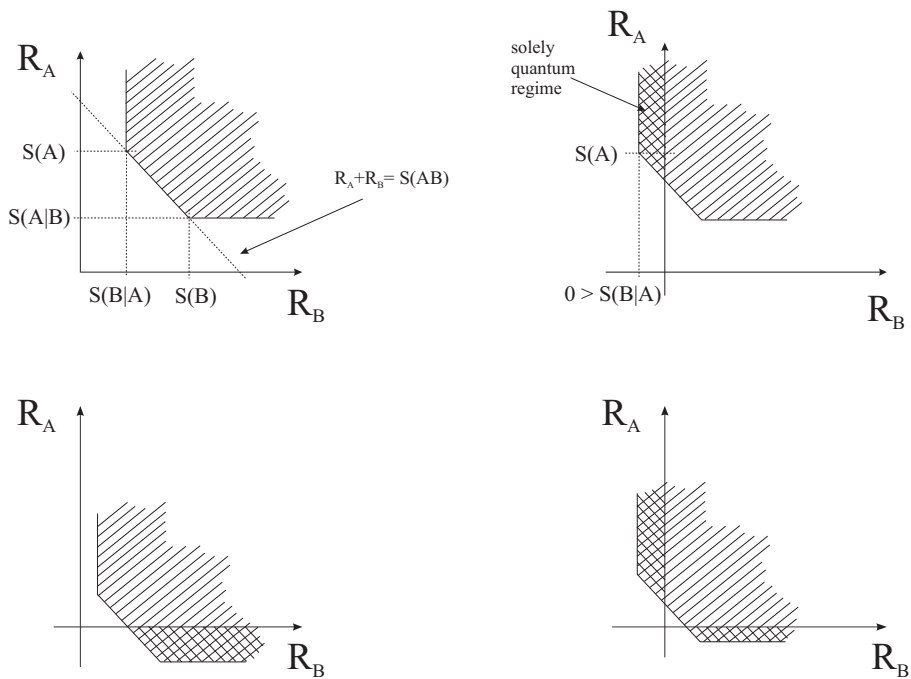}
  \caption{The rate region for distributed compression by two parties
    with individual rates $R_A$ and $R_B$. The total rate $R_{AB}$ is bounded by $\sab$.
    The top left diagram shows the rate region of a source with
    positive conditional entropies; the top right and bottom left
    diagrams show the purely quantum case of sources
    where $\sbca<0$ or $\sacb<0$. It is even possible that
    both $\sbca$ and $\sacb$ are negative, as shown in the
    bottom right diagram, but observe that the rate-sum $S(AB)$
    has to be positive.}
  \label{fig:sw}
\end{figure}

Let us first describe the quantum solution for two parties and depict
the rate region in Figure~\ref{fig:sw}.  
If one party compresses at a rate $\sb$, then the other party can
over-compress at a rate $\sacb$, by merging her state with the state which will end up with
the decoder.  The only difference between this scenario and the state merging one,
is that Bob first compresses his state, and sends it to the decoder, who then decompresses it;
Alice then merges her state with Bob's state which is now at the decoder.  This gives us one
possible way for the two parties to jointly compress the states.
Time-sharing gives the full rate region, since the
bounds evidently cannot be improved.

Analogously, for $m$ parties $A_i$, and all
subsets ${\cal T}\subseteq\{A_1,A_2,\ldots,A_m\}$ holding a combined
state with entropy $S({\cal T})$, the rate sums
$R_{{\cal T}} = \sum_{A_i\in {\cal T}} R_{A_i}$
clearly have to obey
\be
  \label{eq:qSW}
  R_{{\cal T}} \geq S\bigl( {\cal T} | \overline{{\cal T}} \bigr)
                                         \quad \text{ for all sets }{\cal T},
\ee
with $\overline{{\cal T}} = \{A_1,A_2,\ldots,A_m\}\setminus {\cal T}$ the complement
of set ${\cal T}$.
This just follows from the converse to Theorem \ref{thm:merging}: even if the
decoder somehow has all the shares $\overline{{\cal T}}$, a total rate
of at least $S\bigl( {\cal T} | \overline{{\cal T}} \bigr)$ is necessary
to convey the remaining shares ${\cal T}$.

That this bound can be achieved simply follows from the fact that
with $\overline{{\cal T}}$ at the decoder,
each party can in turn merge their state with what will be at the decoder.  So, for 
example, with four parties, an obtainable rate point is obtained when party $A_1$ sends
her state at rate $S(A_1)$ just by regular Schumacher compression, party $A_2$ merges
her state with the first parties state at the decoder with rate $S(A_2|A_1)$, party
$A_3$ merges at a rate $S(A_3|A_1A_2)$, and party $A_4$ at rate $S(A_4|A_1A_2A_3)$,
with rate total being the Schumacher rate $S(A_1A_2A_3A_4)$, etc.
These rate tuples are however just the corners of the region defined
by eq.~(\ref{eq:qSW}); hence time sharing between various
combinations of ordering the encoders gives the full rate region.

\section{Quantum source coding with side information at the decoder}
\label{sec:sideinfo}
Related to distributed compression is the case where only Alice's state needs
to arrive at the  decoder, while Bob can send part of his state to 
the decoder (subject to a rate constraint) in order to help Alice lower her rate.
The classical case of this problem was introduced
by Wyner~\cite{wyner-side}. For the quantum case, we demand that 
the full state $\psi_{ABR}$  be preserved in the protocol, but do not place
any restriction on what part of Bob's state may be at the decoder
and what part can remain with him, while Alice's has to go to the decoder.

To arrive at a formal definition, we would like to speak of
two rates $R_A$ and $R_B$ here, of entanglement between Alice
and the decoder $C$ and of Bob and the decoder $C$. Starting
with $n$ copies of the source, $\Psi_{A^nB^nR^n} = \psi_{ABR}^{\otimes n}$,
we may consider LOCC protocols between $A$, $B$ and $C$,
that take in this state and maximally entangled states
of Schmidt rank $K_A$ ($K_B$) between $A$ and $C$ ($B$ and $C$).
It is supposed to produce a high-fidelity approximation of
$\Psi_{{C'}^nC''B'R^n}$ tensored with maximally entangled states
of Schmidt rank $L_A$ ($L_B$) between $A$ and $C$ ($B$ and $C$),
where $\Psi_{{C'}^nC''B'R^n}$ is obtained from $\psi_{ABR}^{\otimes n}$
by substituting $C^n$ for $A^n$ and with an isometry (e.g. a unitary operation taking one
system to two systems) $B^n \longrightarrow C''B'$.
If in the limit of arbitrary block length the fidelity tends to $1$
and $\frac{1}{n}(\log K_A-\log L_A) \rightarrow R_A$,
$\frac{1}{n}(\log K_B-\log L_B) \rightarrow R_B$, we call the rate pair
$(R_A,R_B)$ achievable, and the side information problem is to
characterise the achievable pairs as concisely as possible.

Using state merging we can see that for any isometry
$T: B \longrightarrow U \otimes V$, the rates 
\begin{equation}
  \label{eq:side-info}
  R_A = S(A|U) \quad \text{ and } \quad R_B = E_P(AU:R)-S(A|U)
\end{equation}
are achievable, where
$\psi_{AUVR} = (\ID_A \otimes T \otimes \ID_R)\psi_{ABR}$, and
\be
  E_P(AU:R) = \min_\Lambda S\bigl( (\ID_{AU}\otimes\Lambda)\rho_{AUV} \bigr)
\ee
is the so-called \emph{entanglement of purification}~\cite{IBMHor2002}
of the state $\rho_{AU\,R}$ with respect to the split $AU$-$R$.
The minimum is taken over all channels $\Lambda$ acting on $V$.
The entanglement of purification is in some sense a measure of total correlations,
as it can be interpreted as the amount of entanglement needed to create a state,
if the only allowed operations is tracing out.

The achievability of rates can be seen as follows: the channel $\Lambda$ can be
represented, with the help of an environment $B'$, as another
isometry $V \longrightarrow WB'$, so that $\psi_{AUVR}$ is mapped
to $\psi_{AUWB'R}$. Now, with many copies, let Bob send the system $U$ to the
decoder, at rate $S(U)$, and Alice merge her state to the
decoder, at rate $R_A = S(A|U)$.
Finally, with the decoder now having $AU$, let Bob merge $W$ to
him, which has rate $S(W|AU)$, so that the total of Bob's
rate is $R_B = S(U) + S(W|AU) = S(AUW) - S(A|U)$. The minimisation
over $W$ leads to the formula for the entanglement of purification.

Here, the isometry $T$ acts on many copies of $B$, and
up to this ``regularisation limit'', the rate pairs (\ref{eq:side-info})
are optimal for one-way protocols.
To see why this is so, consider that at the end of the protocol, 
Bob will have sent part of his state to the decoder.
This part, $U$, is obtained by some
local isometry of Bob's: $B^n \longrightarrow UV$.
Likewise, Alice will have sent all her $A^n$ to the decoder.
The total amount of entanglement used, $n(R_A+R_B)$,
cannot be less than the total
entropy of what ends up at the receiver, which has entropy $S(A^nU)$,
and this is lower bounded by $E_P(A^nU:R)$.
By the converse of Theorem \ref{thm:merging},
Alice's entanglement cost, $nR_A$, cannot be less than $S(A^n|U)$.
Thus we have proved that the set of achievable pairs is given by
\be
  \bigcup_{n=1}^\infty \Bigl\{ \frac{1}{n}\bigl( S(A^n|U),E_P(A^nU:R^n)-S(A^n|U) \bigr)
                            \text{ s.t. } T:B^n\longrightarrow UV \text{ isometry} \Bigr\}.
\ee
(Note that since the formula doesn't mention $V$, we may actually
look at channels $B^n\longrightarrow U$.)

Because $T$ acts on many copies of $B$,
it is unclear whether a single-letter formula for the
achievable rate region can be obtained, potentially by finding
a better -- lower -- expression for Bob's rate.
Indeed, in the classical case, this is what
happens~\cite{wyner-side}. 
For classical random variables $X$ and $Y$ with Alice and 
Bob, respectively, the single-letterized rate for Bob is given by
imagining a channel $Y \rightarrow W$. Bob needs to send only
$I(W:X)$ bits of $W$ rather than $H(W)$.
While the quantum protocol above is clearly optimal, it may be 
that the entanglement of purification is non-additive,
and thus $S(U)$ may be much lower than $nS(U_1)$ where 
$\rho_{U_1}$ is the state obtained by acting a channel
on single copies of $\rho_B$.

\subsection*{Source coding with side information at the encoder}

In the classical case, if a party aims to send her variable 
to the decoder, having herself access to some side information
is of no additional value. If Alice wants to send classical variable 
$X$ to Bob, she cannot lower her rate by
sending or even knowing additional information.
In the quantum world,  this is not the case, as can be seen from
the side information problem in the case of one party.
We consider Alice, who has state $\rho_{A_1}$
and is required to send it to Bob.  This she can do using state
merging at rate $S(A_1|B)$.  However,
if she also has access to state $\rho_{A_2}$ which may be entangled 
or correlated to $\rho_{A_1}$, then she may be able to do better.
This better rate is obtained by sending part of $\rho_{A_2}$ as well --
so in some cases, less is more!

Applying an isometry $T: A_2 \longrightarrow A_2'A_2''$,
and actually merging $A_1A_2'$, she can achieve a rate $S(A_1A_2'|B)$.
Hence one would naturally minimize over channels $T$:
\be
  R \geq \min_T S(A_1A_2'|B).
\ee
As argued in the side information problem, 
the right hand side is equal to $E_P(A_1B:R)-S(B)$.
Essentially, due to the non-monotonicity of the von Neumann entropy, 
it can be beneficial to lower the entropy of what you are sending, 
by merging additional quantum states which
are entangled with what you needed to send.

\section{Multipartite entanglement of assistance}
\label{sec:eofa}
In this section we consider the multipartite entanglement of assistance~\cite{entass}. 
Sometimes it is called localizable entanglement~\cite{vpc2004}, 
although we operate in the regime of many copies and collective measurements. 
Consider a pure $m$-partite state $\psi_{A_1, A_2, \ldots  A_m}$. 
The entanglement of assistance is defined for two fixed nodes $A_i$ and $A_j$,
as the maximal pure entanglement that can be obtained between those nodes 
by LOCC operations performed by all the parties. Here is a more precise definition:

\begin{definition}
For  an $m$-partite  pure state, consider a measurement performed by LOCC  that 
leads to pure states between chosen 
nodes $A_i$ and $A_j$ for any outcome $k$ of the measurement. 
Let the probability of the outcome $k$ be $p_k$,
and the entropy of the node $i$ (equal to entropy of the node $j$) 
be denoted by $S_k(A_i)$.  
The entanglement of assistance between the nodes $A_i$ and $A_j$ is defined as 
\be
  \eass (\psi, A_i:A_j) = \sup \sum_k p_k S_k(A_i) 
\ee
where supremum is taken over the above measurements. 
Asymptotic entanglement of assistance is given by regularization of the 
above quantity 
\be
  \eass^\infty (\psi, A_i:A_j)
           = \lim_{n\rightarrow\infty} \frac{1}{n} \eass(\psi^{\ot n}, A_i:A_j).
\ee
\end{definition}

Asymptotic entanglement of assistance was determined for pure states of up to four 
parties in~\cite{svw2005}. Namely it was proven that for $m\leq 4$
the maximal amount of entanglement that
can be distilled between Alice and Bob, with the help
of the other $m-2$ parties $C_1,\ldots,C_{m-2}$, is
given by the minimum entanglement across any bipartite cut of the system
which separates Alice from Bob:
\beq
  \eass^\infty (\psi, A:B) = \min_{{\cal T}} \{ S(A{\cal T}) , S(B\overline{{\cal T}}) \}
                           =: E_{\text{min-cut}}(\psi, A:B),
\eeq
where the minimum is taken over all possible partitions of the 
other parties into a group ${\cal T}$ and its complement
$\overline{{\cal T}}=\{C_1,\ldots,C_{m-2}\}\setminus {\cal T}$.

In~\cite{SW-nature} we generalized this result to an arbitrary number of parties,
by use of the primitive of state merging. The result is clearly optimal -- one cannot increase
entanglement by LOCC.  The entropy of any splitting ${\cal T}$
which divides $A$ from $B$ is a measure
of the entanglement of the total pure state between $A{\cal T}$ and $B\overline{{\cal T}}$
and it cannot increase during the protocol -- in fact not by any protocol
allowing arbitrary joint operations of the two groups
$A{\cal T}$ and $B\overline{{\cal T}}$ and classical communication.
Thus all entropies under such splitting serve as an upper 
bound for the amount of entanglement which can be distilled between $A$ and $B$.

The protocol for achieving this optimal rate is as follows: 
each party in turn merges their
state with the remaining parties on its side of the minimal cut, 
preserving the minimum cut entanglement.  
The merging protocol we consider will be slightly different from 
the merging protocol considered previously
in two respects.  As before, the party who wishes to merge his 
state with other parties performs
a random measurement on their typical subspace.  However, since the 
receiver will consist of many parties who are separated from one another 
the final decoding step (i.e.~the unitary which the receiver performs 
conditional on the measurement outcome
of the sender) will not be performed until the very end.
The second difference is that the senders will perform complete 
measurements, and will not attempt to 
distill additional entanglement between themselves and the receiving 
parties.  This will not effect the merging condition, but it does
mean that the maximally entangled states which would be created 
between the merging parties and the receiver will
be destroyed.  This greatly simplifies the analysis, despite some 
entanglement being lost.
We only consider entanglement of assistance -- i.e.~a protocol
which attempts to distill entanglement between 
$A$ and $B$.  More complicated protocols
can be constructed which also result in entanglement between other parties.

Before moving to the protocol, we will need to prove an aspect of state merging already
implicit in Theorem \ref{thm:merging}, which will serve as a
cornerstone of (among other things) proving a formula for asymptotic entanglement 
of assistance: for a tripartite pure state $\psi_{ABR}^{\ot n}$ 
if $S(R)<S(B)$, a random rank-$1$ measurement on the typical subspace
$\tA \subset A^n$ produces states $\psi^{j}_{B^nR^n}$ 
such that most of their reduced states $\rho^{j}_{R^n}$ 
are close to the state $\rho_R^{\ot n}$, the reduced state of the
initial state $\psi_{ABR}^{\ot n}$.

\begin{proposition}[Random measurement gives covering]
\label{prop:covering}
Let $\psi_{ABR}$ be a tripartite pure state with $S(R)<S(B)$,
of which we consider $n$ copies, and consider the state $\Psi_{\tA\tB\tR}$ of the
proof of Theorem~\ref{thm:merging}
(Section \ref{sec:main}) belonging to the typical subspaces $\tA\tB\tR$.
Denote by $\rho_{\tR}$ the state of system $\tR$.
Let $\{\ket{e}_j\}$ be a basis on $\tA$ chosen at random according 
to the Haar measure, and $\rho^j_{\tR}$ be the state obtained 
on system $R$ upon obtaining outcome $j$; let $p_j$ 
be the probability of this event.
Then for any $\ep>0$ and all large enough $n$, we have 
\be
  \lavv \sum_j p_j \bigl\| \rho^j_{\tR} - \rho_{\tR} \bigr\|_1 \ravv  \leq \ep,
\label{eq:covering}
\ee
where the average is taken over the choice of basis.
\end{proposition}

\begin{beweis}
  This is just the special case of $L=1$ in Proposition~\ref{prop:one-shot:merge}.
\end{beweis}

With this tool in hand we can analyze the protocol outline above.
Clearly, if $m=2$, there is only one cut, and its entropy is $S(A)$,
the entropy of entanglement, and we are done. So, from now on $m\geq 3$.

Assume for the moment that all $S(A{\cal T})$ are distinct (we'll come
back to this point at the end), and consider helper $C_{m-2}$.
For each set ${\cal T}$, clearly $S(A{\cal T}) = S(B\overline{{\cal T}})$,
by the purity of the overall state. Hence, for the min-cut we can
restrict to looking at the entropies $S(A{\cal T})$ and $S(B{\cal T})$,
with $C_{m-2} \not\in {\cal T}$. For each such set
${\cal T} \subset \{ 1,\ldots,m-3 \}$, consider the relative
complement ${\cal T}' := \{ 1,\ldots,m-3 \} \setminus {\cal T}$.
This defines a tripartite system composed of
$C_{m-2}$, $A{\cal T}$ and $B{\cal T}'$. Let $C_{m-2}$ perform a random
measurement on his typical subspace $\widetilde{C}_{m-2}$, as in
Proposition~\ref{prop:covering}. We get (if only $n$ is large enough),
with arbitrarily high probability,
states $\Psi^j_{ABC_1\ldots C_{m-3}}$ which by eq.~(\ref{eq:covering})
satisfy:
\be
  \text{For all } {\cal T}: \quad
  S\bigl(A^n{\cal T}^n\bigr)_{\Psi^j} = S\bigl(B^n{{\cal T}'}^n\bigr)_{\Psi^j}
                = n\Bigl( \min\bigl\{ S(A{\cal T}), S(B{\cal T}') \bigr\} \pm \delta \Bigr),
\ee
with arbitrarily small $\delta$. In other words, for each such $\Psi^j$,
\be
  E_{\text{min-cut}}\bigl(\Psi^j,A^n:B^n\bigr)
                   = n\bigl( E_{\text{min-cut}}(\psi,A:B) \pm \delta \bigr),
\ee
and that means that the min-cut entanglement is almost preserved (up to
an arbitrarily small variation in the rate), and hence that the reduced
state entropies can be assumed to be all distinct (by choosing $\delta$
small enough). Now we recursively apply the same to $C_{m-3},\ldots,C_1$.
\hfill $\Box$

Finally, for the assumption that all reduced state entropies are pairwise distinct:
this can be enforced if the parties first ``borrow'' an arbitrarily small
rate of entanglement to distribute singlets between chosen pairs.
Then our distinctness assumption becomes true.
In the limit, only a sublinear amount of entanglement is needed to
do this, but on the other hand~\cite{SmolinThapliyal:little-o} shows that the asymptotic
entanglement landscape of multiple parties does not change if
one allows this sublinear amount -- this is due to them being able
to always, perhaps inefficiently, extract \emph{some} entanglement
across any given cut unless across that cut they happen to be in
a product state.

\begin{remark}
Note that a crucial part of the argument of why the minimum cut entropy doesn't change
is the use of random codes.  This is because $C_1$'s procedure
is universal -- it does not depend on the cut.  He makes a measurement
which only depends on the
typical subspace of his state.  The measurement thus serves to
merge his state with whichever
grouping of subsystems has the larger entropy compared
with the remaining systems.  Not all
quantum codes have this feature -- for example Devetak
codes~\cite{igor-cap} depend both on
the state of the sender, and that of the receiver.
The same applies to~\cite{svw2005}, which is why there even the argument
for $m=4$ has to be quite subtle.
\end{remark}

It may seem odd that after performing a random measurement,
ones state goes to any set of parties which has more entropy than the remaining
parties.  Since there are many possible groupings of the parties, 
for some groupings a certain party would help receive the state, but for other groupings,
that party's state would be left unchanged by the random measurement.  Of course, there
is no contradiction, as in the end, at the decoding step,
one has to decide on the grouping, and with
fidelity approaching $1$ only for many copies of the state.

\begin{conjecture}
  \label{conj:simultaneous}
  It is awkward that in the recursive procedure described above for
  $m$ parties we have to first consider a measurement on a long block
  of states, and then for the second measurement blocks of these blocks, etc.

  It seems likely that the simplest random measurement strategy will
  indeed also work: all $m-2$ helpers $C_1,\ldots,C_{m-2}$ measure
  in a random basis of their respective typical subspaces and broadcast
  the result to Alice and Bob. They should then end up, with high probability,
  with a state of the min-cut entanglement.
\end{conjecture}

\section{Capacity region for the multiple access channel}
\label{sec:multiply-multiple}
We consider a channel with two senders Alice and Bob, 
and one receiver Charlie; this is the multiple access channel. 
For the classical multiple access channel, any rates satisfying
the following inequalities are achievable for encoding independent
messages from Alice and from Bob at their respective terminals
to Charlie who decodes them jointly:
\begin{align}
  R_A       &\leq I(A:C|B) \nonumber\\
  R_B       &\leq I(B:C|A) \label{eq:classical-mac} \\
  R_A + R_B &\leq I(AB:C). \nonumber
\end{align}
The quantum multiple access channel -- where Alice and Bob want to send quantum information
was considered in~\cite{ydh2005}, and we refer to that paper for
the definitions of codes and rate region.
In~\cite{SW-nature}, we found that one could use state merging to find
a larger achievable region, including
negative rates. Namely, that for the quantum multiple access channel, 
there is the following region of achievable rates:
\begin{align}
 R_A      &\leq I(A\>C|B) := I(A\> BC) \nonumber \\
 R_B      &\leq I(B\>C|A) := I(B\> AC) \label{eq:rates1} \\
 R_A+ R_B &\leq I(AB\>C).              \nonumber
\end{align}
The state on which the quantities are evaluated is constructed as follows.
Consider two pure states 
$\psi_{AA'}$ and $\psi_{BB'}$. Let $\rho_{ABC}$ be the state, resulting 
from the halves $A'$ and $B'$ being sent down the channel:
\be
\rho_{ABC}= (I_{AB} \ot \Lambda_{A'B'\to C}) (\proj{\psi}_{AA'} \ot \proj{\psi}_{BB'}).
\ee
In the classical theory, only positive rates make sense.
In the quantum case, the rates can be  meaningful, even if one of them is negative. 
For example, when $R_A$ is negative, and $R_B$ is positive,
this means that when Alice invests $R_A$ qubits, then Bob can send $R_B$ qubits,
as we shall see.

\subsection{Remarks on coherent information}
In~\cite{Schumacher-Nielsen} the coherent information was introduced and defined 
in terms of an input state $\rho_A$ and a channel producing output $\rho_B$ as 
\be
  I(A\>B) = S(B) - S(AB),
\ee
that is, as the conditional entropy with a minus sign; this was puzzling
because it can be negative.  Since it gives the channel capacity of a quantum
channel (by maximizing it over input distributions $\rho_A$), it was unclear
how to interpret negative uses of a channel.
We will see that the negative part will acquire operational meaning,
in full accordance with the positive part.
We also have defined the conditional coherent information as
\be
  I(A\> B|C)=S(B|C) - S(AB|C).
\ee 
We have the useful identity [consistent with eq.~(\ref{eq:rates1})]
\be
  I(A_1 \> B| A_2)= I(A_1 \> B A_2).
\ee
That is, conditioning the coherent information is very simple: just erase the bar. 
Then we have a chain rule of the same form as the one for mutual information,
\be
  I(A_1 A_2 \> B)= I(A_2 \> B) + I (A_1\> B |A_2).
  \label{eq:chain}
\ee
What seems surprising 
is that conditioning can only increase coherent information!
However, this can be explained as follows. Namely, in classical information theory
we have to have situations where conditioning decreases information, 
due to \emph{lack of monogamy}. Indeed, we can have situation where
\be
  I(X_1:Y) + I(X_2:Y) > I(X_1 X_2:Y).
\ee
(E.g., the three variables could be fully correlated.)
Therefore, to save the chain rule, conditioning must decrease 
mutual information. 
However in the quantum case we always have 
\be
  I(A_1 \> B) +I(A_2 \> B) \leq I(A_1A_2 \> B),
\ee
due to strong subadditivity.  Now conditioning very often increases 
coherent information, because we have equality in the chain rule
identity (\ref{eq:chain}).

\subsection{Direct coding theorem: achievability of rates}
\begin{figure}[ht]
  \includegraphics[width=12cm]{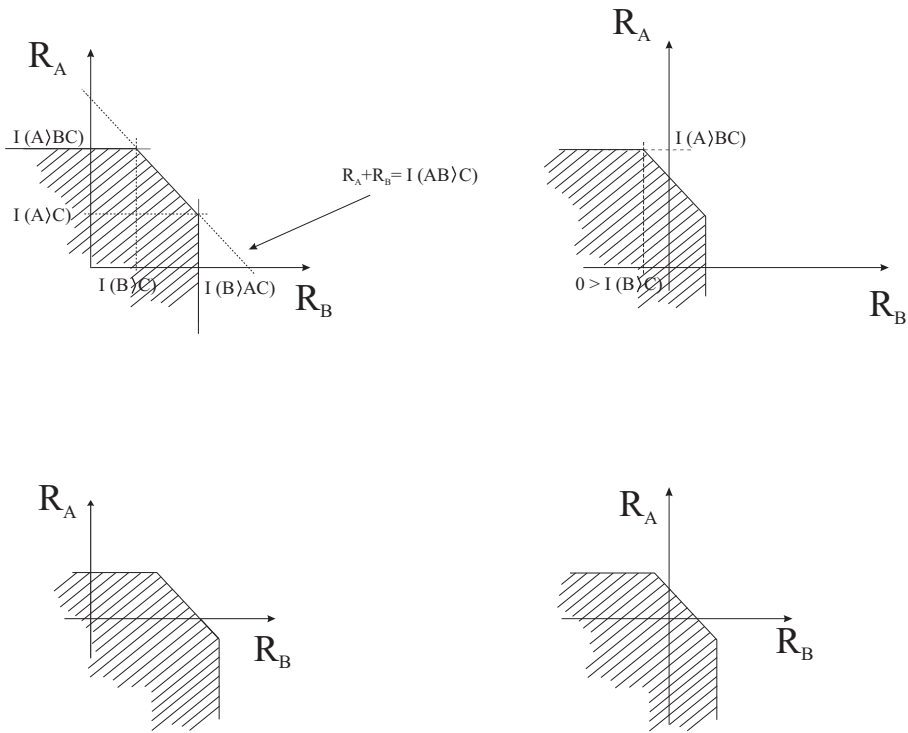}
  \caption{
The rate region for the multiple-access channel for two parties
    with individual rates $R_A$ and $R_B$. The total rate $R_{AB}$ is 
bounded by $I(AB\>C)$.
    The top left diagram shows the rate region when both rates are 
    positive; the top right and bottom left
    diagrams show the case
    where $I(B\>C) < 0$ or $I(A\>C)<0$. I.e. here, Bob (Alice) can
invest entanglement so that the other party can send at a rate  
$I(A\>BC) \geq R_A \geq I(A\>C)$ ($I(B\>AC) \geq R_B \geq I(B\>C)$). 
    In the bottom right diagram, both parties may have the option of
achieving the higher rate by having the other party invest entanglement.
}
  \label{fig:channel-rates}
\end{figure}

To check that the rates satisfying the above conditions are 
achievable, it is enough to consider one corner, for example
\be
  R_A = I(A\>BC),\quad R_B = I(B\>C),
\ee
which is an upper corner of the rate region, see Figure \ref{fig:channel-rates}.

When both $I(A\>BC)$ and $I(B\>C)$ are negative, they are trivially 
achievable: Alice and Bob do nothing. 
So in this case negativity of rates does not appear meaningful,  
as zero is achievable too, and one always optimises rates over input states. 
When $I(A\>BC) $ is negative and $I(B\>C)$ 
is positive, again, those rates can be achieved by Alice doing nothing, 
and Bob -- by standard quantum coding theorem. So again the negative rate is 
not interesting. There are therefore two situations, which we have to consider:
\begin{align}
  \label{eq:case1}
  I(A\>BC) &\geq 0 \quad\text{and}\quad I(B\>C) \geq 0, \quad\text{or} \\
  \label{eq:case2}
  I(A\>BC) &\geq 0 \quad\text{and}\quad I(B\>C) < 0.
\end{align}
It is enough to consider the first one in detail, as the second one is its 
simple consequence. Let us first describe how to achieve those rates, 
when Bob and Alice can communicate quantum messages to $C$ if classical
side-communication is permitted.
Alice and Bob prepare ($n$ copies of) states $\psi_{AA'}$ and $\psi_{BB'}$,
respectively, and send halves of them down the channel
(inputs ${A'}^n$ and ${B'}^n$).
Then Bob performs the merging protocol, 
i.e.~he makes the measurement on his typical subspace in blocks 
of size $2^{nR_B}$. As previously we label blocks (codes) by $j$. 
On average, he obtains a state close to a $2^{nR_B}$ 
dimensional maximally entangled state shared with Charlie (who holds the system C),
and Bob's part of the state $\psi_{ABCR}$ is merged with Charlie 
($\psi_{ABCR}$ is purification of $\rho_{ABC}$). 
Then, Alice shares with Charlie state $\rho_{ABC}$ 
where both part $B$ and $C$ is now with Charlie. 
Random measurement of Alice in blocks $2^{nR_A}$, 
will create a state close to the maximally entangled state 
of this dimension between Alice and Charlie,
after Alice communicates her results to Charlie. In this way she also 
merges her part to Charlie,  however it is not important in the present context.

Let us now show, how Alice and Bob can share with Charlie 
maximally entangled state of suitable dimensions without classical communication. 
Namely, both Alice and Bob can perform their measurements before 
sending halves of their states $\psi_{AA'}$ and $\psi_{BB'}$ 
down the channel. They can then send the states $\psi_{AA'}^{j_A}$,
 $\psi_{BB'}^{j_B}$ that 
they have  obtained (here $j_A$ and $j_B$ denote the outcomes of measurement).
This still requires communication, 
as they have to tell Charlie, what outcomes they obtained. 

However, instead of measuring, they can prepare already 
$\psi_{AA'}^{j_A}$,  $\psi_{BB'}^{j_B}$ with fixed $j_A$ and $j_B$ 
known to Charlie. This will have the same effect as before, once they choose such labels, 
that guarantee that merging conditions are satisfied. 
Note that the states that Alice and Bob are now sending are close to maximally entangled states
(this is guaranteed by the merging condition). 
The maximally entangled states to which they are close, defines the subspaces, 
which go through the channel, and allow correction of errors. 
The subspaces are codes that when used by Alice and Bob,
allow them to obtain the above rates.  Since our criterion 
was fidelity with the maximally entangled state, we have obtained here 
coding theorem with small average error. 

In our case it was relatively easy to go from one way to zero 
because the states that Alice and Bob obtain in our one-way protocol 
are close to maximally entangled states. For more complicated situations 
see~\cite{DemianowiczH2005}. 

Finally, consider the case, where $I(B\>C)$ is negative,
eq.~(\ref{eq:case2}). The reasoning is very similar:
in the scenario with classical side-communication, 
Bob sends $-I(B\>C)+\ep$ halves of maximally entangled states through a noiseless channel,
(keeping the other half), and performs 
merging, so that after that Alice can achieve her rate as above.
However, again Alice and Bob instead of performing measurements,
can send the state that would emerge under some outcome of 
the measurement. The difference is that Bob will 
send the state not only down the noisy channel, 
but also down the supplementary noiseless channel, and will 
share $\ep$ rate of maximally entangled states (thus his overall rate 
is negative). This is the more interesting rate point: for Alice to achieve
the rate $I(A\>BC)$, she requires Charlie to have $C$ and $B$.
Bob assists in providing this information (which can be understood as
additional error correcting information from inside the channel)
but that comes at a price, which is exactly $-I(B\>C)$.  We thus have an interpretation
of negative channel capacities.   \hfill $\Box$

\subsection{Converse coding theorem}
Here we briefly argue that (up to regularization)
the rate region described by our conditions
is optimal. The reasoning is quite standard
(see e.g.~\cite{BarnumNS-converse1997,HHH-hashing}), therefore
we will provide only a sketch of the proof.
Suppose that some rates $R_A$ and $R_B$ are achievable.
Consider first the case where they are both positive.
This means that Alice and Bob can send halves of singlets
down the channels in such a way
that after decoding by Charlie, they share with Charlie
those singlets with fidelity tending asymptotically to one.
Alice shares a singlet of dimension $2^{n R_A}$ with Charlie,
and Bob one of dimension $2^{nR_B}$. Would they have exact singlets,
the coherent informations would be equal to $I(A\>BC)=I(A\>C)=n R_A$,
$I(B\>C)=nR_B$ and $I(AB\>C)=n(R_A+R_B)$.
Because they share inexact singlets, we apply asymptotic continuity
of coherent information~\cite{HHH-hashing} (which plays here the role of Fano's
inequality),
thanks to which the coherent informations of the real state,
per use of channel, approach the ideal values in the asymptotic limit.
This means that there exist such states,
such that, if Alice and Bob will send halves of them down the channel,
then after Charlie's decoding, the coherent informations
approach the values from the coding theorem.

There are still two issues. First, the states may be mixed:
Alice and Bob prepared singlets, however the encoding procedure may
turn them into mixed states. However, coherent information is convex,
so that Alice and Bob will not do worse by sending some pure states.
Second, we considered the joint $ABC$ state after Charlie's decoding,
while in the coding theorem, we have state merging just from sending
by Alice and Bob. However,
due to the data processing inequality~\cite{BarnumNS-converse1997} (saying that
operating on $V$ one cannot increase $I(U\>V$), the coherent information
of the state before Charlie's decoding can be only greater.

Let us now consider the case when one of the rates (suppose $R_B$) is negative.
This means that Bob uses the noiseless qubit channel an additional
$R_B$ times (per use of the noisy channel), 
and Alice achieves her rate.
It suffices to show that, if rate pair $(R_A,R_B)$
where $R_B$ is negative, is achievable,
then Alice and Bob can create the joint state of $ABC$ system,
such that $I(A\>BC)=R_A$ and $I(A\>C)=R_B$ per use of channel.
To this end, consider a new channel which consists of the old one
supplemented by $-R_B+\ep$
uses of the noiseless channel from Bob to Charlie. For the new channel, the
rates $(R_A, \ep)$ are achievable. They are positive, so that, as explained
above, there exist states of Alice and Bob, that sent down the channel
produce a joint state having $I(A\>C)=R_A$ and $I(B\>C)=\ep$.
Suppose now that Bob will not send part the system that
was intended to go through the noiseless channel, but keeps it.
In this situation they only use the original channel.
We will now see that they achieve the needed coherent informations
in this way. Of course $I(A\>BC)=R_A$, as this quantity does not depend
on whether a given system is with Bob or with Charlie. Let us now estimate
the quantity $I(B\>C)$. By sending $-R_B+\ep$ qubits, Bob could increase it
up to $\ep$. However, by sending one qubit, one can increase coherent
information
no more than by one. Thus, coherent information $I(B\>C)$ cannot be smaller
than
$R_B$. This ends the proof of the converse theorem.

\section{Strong subadditivity}
\label{sec:strongsa}
Using state merging, we can get a very quick and operationally intuitive  proof
of strong subadditivity~\cite{lieb:ruskai}, which can be written as
\beq
  S(A|BC)\leq S(A|B).
\eeq
Strong subadditivity is simply the observation that if Bob has access to
an additional register $C$, then Alice surely doesn't need to send more
partial information for him to get the full state $\rho_{AB}$.  After all,
Bob could always ignore the ancilla on $C$, but if he uses it, Alice may need
to send him less. Mathematically, we can use this argument because
in the proof that $S(A|B)$ is the optimal merging rate we have used only
typical subspaces and elementary probability for the direct part,
and ordinary subadditivity in the converse part.

\section{Conclusion}
\label{sec:conclusion}
It is very interesting to compare the proof of the classical Slepian-Wolf
theorem, with the proof of its quantum version -- state merging. 
The Slepian-Wolf protocol is as follows: the typical sequences of Alice 
are divided into blocks of size $\approx 2^{n I(A:B)}$. 
Note that this is the size of a good code. 
Now, when a particular sequence occurs, 
Alice lets Bob know in which code is the sequence, 
and this is enough for him to determine her sequence. 
Thus the Slepian-Wolf theorem follows solely from the fact that a random code is a 
good code, which was shown by Shannon.

Interestingly, our protocol is based on the same property,
especially for states for which coherent information is positive. 
(This could be regarded as a situation analogous to the classical case,
as the classical mutual information is always positive.)
Namely, to prove quantum state merging it is enough to know that a random 
quantum code is a good quantum code. 
And in the quantum state merging protocol  Alice performs an analogous task:
she measures in which quantum code her state is, and tells Bob the result.

What is now  extremely surprising, is that those similarities 
turn out to be quite superficial. Namely, 
in the Slepian-Wolf protocol, the amount of bits needed to tell 
Bob the information ``which code'' is just the cost of transmission 
of Alice's data to Bob. 
In the quantum case, the information ``which code'', since represented by classical 
bits, is not counted at all, as we count only the quantum information.
Thus in this case (positive coherent information) merging 
does not cost at all, unlike in the classical case.
What is more remarkable still is that despite this difference, the cost 
of sending partial quantum information is the conditional entropy, and
thus formally similar to the classical case.  This despite the fact that
the classical case does not emerge as a limit from the quantum case.  In
other words, if one takes quantum state merging, and applies it to classical
states (i.e.~states which are fully decohered, and contain only classical correlations),
then the goal is rather different, as one is attempting to retain entanglement 
between this classical state and the reference system, and one is further allowing
free classical communication.

We have {\it two ways} of interpreting the classical mutual information:
(i) either as the quantity responsible for capacity 
or (ii) as the quantity that reports the part of information that is common both to Alice 
and Bob. Indeed, the latter meaning is implied by the fact that 
the cost of communication needed to transfer full information 
to Bob is $H(X)$ (full information content of Alice's state) 
reduced by the amount of mutual information. Thus the latter 
represents  that part of Alice's information,
that Bob also knows, and it need not be transferred to him. 

It turns out that in the quantum case  those two notions 
are no longer represented by the same quantity (see however~\cite{BSST-ea}).
Namely, the communication cost is equal to 
Alice's information reduced by quantum mutual information.
Thus quantum mutual information serves as common information.
The capacity is on the other hand represented by the coherent information. 
The first quantity is sometimes greater than the whole of Alice's information,
and precisely in those instances, the second quantity has the chance to be 
positive. 

It is indeed the beauty of the quantum information world, 
that both the quantities, into which the classical quantity has split, 
do their job in an analogous way as it was in the classical case.
Indeed, the analogue of common information counts by how much the
transmission cost is reduced -- exactly as in the classical case, 
while the analogue of capacity is responsible for protocol,
with the same basic elements as in classical case. 
The additional brick in the quantum protocol is teleportation,
which is perhaps the thread that binds the two notions together. 

However, as we have noted, the analogy in the protocol is quite superficial. 
Even though Alice perform the operations that can be called 
by use of the same name (checking ``which code'', and telling it to Bob)
the meaning of those operations is completely different. 
It is extremely mysterious, how the quantum and classical cases 
can have so much in common, and at the same time can be so different.

\acknowledgments
MH acknowledges EC grants RESQ, QUPRODIS, EC IP SCALA
and (solicited) grant of Polish Ministry of Science and Education
contract no.~PBZ-Min-008/P03/03,
JO acknowledges the support of
the Royal Society, the Cambridge-MIT Institute, and EU grant
PROSECCO. 
AW thanks the EC for support through the RESQ project, 
the U.K. EPSRC for support through the ``QIP IRC'', and the University
of Bristol for a Research Fellowship.

\appendix

\section{Miscellaneous facts about norms and fidelity}
\label{app:misc}
The following lemma relates the \emph{trace norm} to the
\emph{Hilbert-Schmidt norm}.
Recall that these norms are defined, for an operator $X$, as
\begin{align}
  \| X \|_1 &:= \tr\sqrt{X^\dagger X}, \tag{\text{trace norm}}\\
  \| X \|_2 &:= \sqrt{\tr X^\dagger X} \tag{\text{Hilbert-Schmidt norm}}.
\end{align}

\begin{lemma}
\label{lem:norms}
For any operator $X$,
\be
  \|X\|_1^2 \leq d \|X\|_2^2 ,
\ee
where $d$ is the dimension of the support of operator $X$ (the subspace on which 
$X$ has nonzero eigenvalues).
\end{lemma}
\begin{beweis}
  It is implied by convexity of function $x^2$, where 
  one takes probabilities $1/d$.
\end{beweis}

The \emph{fidelity} of two states is given by 
\be
  F(\rho,\sigma) = \left( \tr \sqrt{\sqrt \rho \sigma \sqrt \rho} \right)^2.
\ee
Notice that if one of the states is pure, say $\sigma=\proj{\phi}$, then
\be
  F(\rho,\proj{\phi}) = \bra{\phi}\rho\ket{\phi} = \tr(\rho\proj{\phi}).
\ee

\begin{lemma}
\label{lemma:fid-norm}
The fidelity is related to trace norm as follows~\cite{Fuchs-Graaf}:
\be
  1-\sqrt{F(\rho,\sigma)} \leq \frac{1}{2} \|\rho-\sigma\|_1 \leq \sqrt{1-F(\rho,\sigma)}.
  \label{eq:fuchs}
\ee
\end{lemma}

\begin{lemma}[Gentle measurement]
\label{lemma:gentle}
Let $\rho$ be a (subnormalized) state, i.e.~$\rho\geq 0$  and $\tr \rho\leq 1$,
and let $0\leq X\leq I$. Then, if $\tr \rho X \geq 1- \ep$,
\be
  \bigl\| \sqrt{X} \rho \sqrt{X} - \rho \bigr\|_1 \leq 2\sqrt{\ep}.
\ee
\end{lemma}
\begin{beweis}
  See~\cite{winter:strong}, Lemma 9; the better constant above is
  from~\cite{OgawaNagaoka2001}.
\end{beweis}

\begin{lemma}[Fannes~\cite{Fannes1973}]
  \label{lem:fannes}
  For states $\rho$ and $\sigma$ on a $d$-dimensional space, such that
  $\| \rho-\sigma \|_1 \leq \epsilon$,
  \be
    \bigl| S(\rho)-S(\sigma) \bigr| \leq \eta(\epsilon)\log d,
    \quad
    \text{with}
    \quad
    \eta(x) := \begin{cases}
                 x-x\log x          & \text{ if }x\leq\frac{1}{e}, \\
                 x+\frac{\log e}{e} & \text{ if }x\geq\frac{1}{e}.
               \end{cases}
  \ee
\end{lemma}

\section{The twirling average of eq.~(\ref{eq:twirled})}
\label{app:twirl}
We use the fact that an operator
\be
  {\cal T}(X) := \bigl\langle ({UU}_{\tA\tA})^\dagger X ({UU}_{\tA\tA}) \bigr\rangle
\ee
is $U\ot U$-invariant. However, the representation of $U\ot U$
decomposes into the two irreducible components, the
symmetric and the antisymmetric subspace. By Schur's lemma,
the only invariant operators are then linear combinations
of the projections onto these subspaces:
\be
  \Pi_{\tA\tA}^{\rm sym}  = \frac{1}{2}\bigl( I_{\tA\tA} + F_{\tA\tA} \bigr), \qquad
  \Pi_{\tA\tA}^{\rm anti} = \frac{1}{2}\bigl( I_{\tA\tA} - F_{\tA\tA} \bigr).
\ee
Hence, the twirling map ${\cal T}$ can be written
\be
  {\cal T}(X) = \frac{1}{\tr\Pi_{\tA\tA}^{\rm sym}} \Pi_{\tA\tA}^{\rm sym}
                                     \tr\bigl( X\Pi_{\tA\tA}^{\rm sym} \bigr)
                + \frac{1}{\tr\Pi_{\tA\tA}^{\rm anti}} \Pi_{\tA\tA}^{\rm anti}
                                     \tr\bigl( X\Pi_{\tA\tA}^{\rm anti} \bigr).
\ee
This is enough to evaluate our average:
\begin{equation}\begin{split}
  \bigl\langle ({UU}_{\tA\tA})^\dagger F_{A_1A_1} ({UU}_{\tA\tA}) \bigr\rangle
    &= \frac{2}{d_{\tA}(d_{\tA}+1)} \Pi_{\tA\tA}^{\rm sym}
                                    \tr\bigl( F_{A_1A_1}\Pi_{\tA\tA}^{\rm sym} \bigr)
      + \frac{2}{d_{\tA}(d_{\tA}-1)} \Pi_{\tA\tA}^{\rm anti}
                                     \tr\bigl( F_{A_1A_1}\Pi_{\tA\tA}^{\rm anti} \bigr) \\
    &= \frac{2}{d_{\tA}(d_{\tA}+1)} \Pi_{\tA\tA}^{\rm sym} \frac{L+L^2}{2}
      + \frac{2}{d_{\tA}(d_{\tA}-1)} \Pi_{\tA\tA}^{\rm anti} \frac{L-L^2}{2}            \\
    &= \frac{L(L+1)}{d_{\tA}(d_{\tA}+1)} \frac{I_{\tA\tA}+F_{\tA\tA}}{2}
      - \frac{L(L-1)}{d_{\tA}(d_{\tA}-1)} \frac{I_{\tA\tA}-F_{\tA\tA}}{2}               \\
    &= \frac{L}{d_{\tA}}\frac{d_{\tA}-L}{d_{\tA}^2-1}I_{\tA\tA}
       + \frac{L}{d_{\tA}}\frac{Ld_{\tA}-1}{d_{\tA}^2-1}F_{\tA\tA}.
\end{split}\end{equation}

\section{Typicality}
\label{app:typicality}
We shall need the concept and a few properties of typical subspaces
\cite{Schumacher1995}. Consider $n$ copies of a density matrix $\rho$,
$\rho^{\ot n}$. Writing $\rho$ in its eigenbasis,
$\rho = \sum_i p_i \proj{i}$, we note first of all that
$S(\rho) = H(p_i)$.
Now,
\be
  \rho^{\ot n} = \sum_{i^n} p_{i^n} \proj{i^n},
\ee
with
\begin{align}
  i^n       &= i_1\ldots i_n,            \nonumber \\
  p_{i^n}   &= p_{i_1}\cdots p_{i_n},              \\
  \ket{i^n} &= \ket{i_1}\cdots\ket{i_n}. \nonumber
\end{align}
For $\delta>0$, the set of \emph{typical sequences} is defined
as (see~\cite{CoverThomas})
\be
  {\cal T}^n_\delta := \bigl\{ i^n : |-\log p_{i^n} - nS(\rho)| \leq n\delta \bigr\},
\ee
and the \emph{typical projector}~\cite{Schumacher1995} is
\be
  \Pi^n_\delta := \sum_{i^n\in{\cal T}^n_\delta} \proj{i^n}.
\ee

The typical projector inherits its properties from the set of
typical sequences. We quote the following from~\cite{Schumacher1995},
and from~\cite{Winter1999PhD} for the exponential bounds
(see also~\cite{CoverThomas}):
abbreviating $\Pi=\Pi^n_\delta$,
\begin{align}
  \label{eq:C1}
  \tr(\rho^{\ot n}\Pi) &\geq 1-\exp(-c\delta^2 n) \ \text{ with a constant }c, \\
  \label{eq:C2}
  \Pi\rho^{\ot n}\Pi   &\leq \rho^{\ot n},                                   \\
  \label{eq:C3}
  \Pi\rho^{\ot n}\Pi   &\leq 2^{-n[S(\rho)-\delta]}\Pi,                     \\
  \label{eq:C4}
  \Pi\rho^{\ot n}\Pi   &\geq 2^{-n[S(\rho)+\delta]}\Pi,                    \\
  \label{eq:C5}
  \rk\Pi =    \tr\Pi   &\leq 2^{n[S(\rho)+\delta]},                       \\
  \label{eq:C6}
  \rk\Pi =    \tr\Pi   &\geq \bigl(1-e^{-c\delta^2 n}\bigr)
                                                   2^{n[S(\rho)-\delta]}.
\end{align}



\end{document}